\documentclass[letterpaper,11pt]{article}
\usepackage[margin=1in]{geometry}
\usepackage[T1]{fontenc}
\usepackage{lmodern,microtype}
\usepackage{amsmath,amssymb,amsthm,mathtools}
\usepackage{enumitem,xspace,booktabs}
\usepackage{algorithm}
\usepackage[noend]{algpseudocode}
\usepackage[authoryear,round]{natbib}
\usepackage{xcolor}
\usepackage[colorlinks=true,linkcolor=blue,citecolor=blue,urlcolor=blue]{hyperref}
\usepackage[capitalize,nameinlink]{cleveref}
\makeatletter
\providecommand{\theHALG@line}{\thealgorithm.\arabic{ALG@line}}
\makeatother
\setlist{itemsep=3pt,topsep=5pt}
\allowdisplaybreaks
\newtheorem{theorem}{Theorem}
\newtheorem{lemma}{Lemma}
\newtheorem{proposition}[theorem]{Proposition}

\theoremstyle{definition}
\newtheorem{definition}{Definition}
\newtheorem{example}{Example}
\theoremstyle{remark}
\newtheorem{remark}{Remark}
\newcommand{\R}{\mathbb R}

\newcommand{\eps}{\varepsilon}
\newcommand{\sm}{\sum}
\newcommand{\mpb}{\textup{MPB}\xspace}

\newcommand{\KL}{D_{\mathrm{KL}}}
\DeclareMathOperator*{\argmin}{arg\,min}
\DeclareMathOperator*{\argmax}{arg\,max}

\title{A Polynomial-Time Algorithm for \\Constant-Factor EFX Allocations of Chores}
\author{Aniket Murhekar \\
\small{\texttt{aniket1602@gmail.com}}}
\date{\small Northwestern University}
\begin{document}
\maketitle
\begin{abstract}

We study the fair division of indivisible chores among agents
with additive disutilities, focusing on the central fairness notion of envy-freeness up
to any chore (EFX), and its multiplicative approximations. Although \citet*{GMQ25} and \citet{GM26}
established the existence of $4$-EFX and $2$-EFX allocations,
respectively, computing a constant-factor EFX allocation in
polynomial time remained open. We resolve this question by
giving an algorithm that, for every $\eps\in(0,1)$, computes a
$(3+\eps)$-EFX allocation in time polynomial in the input length
and $1/\eps$.

Our result relies on two algorithmic contributions. First,
we give a fully polynomial-time approximation scheme (FPTAS) for
earning-restricted (ER) equilibria with equal earning requirements
and arbitrary feasible chore earning limits. We develop a
new nonconvex program with chore payouts and money-flow variables, and prove that every local maximum yields an exact ER
equilibrium. We then give a mirror-ascent algorithm for this program
and prove that it converges to an approximate ER equilibrium in
polynomial time.

Second, we give a new procedure that converts the 
approximate equilibrium into a $(3+\eps)$-EFX-friendly allocation,
a suitable starting point for the chore-swap framework of
\citet{GM26}. The procedure uses a flexible price threshold to
choose swap chores and their recipients together, overcoming a
factor-four bottleneck in the earlier approaches while
accommodating approximate earnings. We expect these ideas to be
useful more broadly in fair division, including for goods and
other fairness notions.

\end{abstract}

\section{Introduction}\label{sec:intro}

Fair allocation is a fundamental problem studied across computer science,
economics, operations research, and mathematics. It is concerned with
allocating resources or responsibilities to agents with heterogeneous
preferences in a fair manner. In this paper, we study the fair division
of \textit{indivisible chores}, which impose a disutility or cost on the
agents receiving them. Examples include assigning household tasks,
work shifts, and maintenance responsibilities. We assume agents have
\textit{additive disutilities}: every chore imposes a positive
cost, and the disutility an agent incurs from a set
of chores is the sum of her disutilities for its individual chores.

\textit{Envy-freeness} (EF) is one of the most fundamental notions of
fairness~\citep{foleyEF}. An allocation is envy-free if every agent
prefers her own bundle to the bundle assigned to any other agent.
Unfortunately, EF allocations need not exist when chores are indivisible,
as can be seen by considering a single chore and two agents. This
motivates relaxations of EF for the discrete setting. 
A compelling relaxation is \textit{envy-freeness up to any chore}
(EFX), which requires that every agent
prefers her own bundle to any other agent's bundle after the removal of
\textit{any} chore from her own bundle. Formally, an allocation
$X=(X_1,\ldots,X_n)$ is EFX if
$d_i(X_i\setminus\{j\})\le d_i(X_h)$ for all agents $i,h$ and
$j\in X_i$; here $d_i$ denotes agent $i$'s additive disutility function and $n$ is the number of agents.

EFX allocations of chores need not exist even for additive
disutilities~\citep{HeTao2026}. To provide meaningful fairness guarantees, it is therefore natural and important to investigate
\textit{multiplicative approximations} to EFX. In a $\lambda$-EFX
allocation, where $\lambda\ge1$, the above inequality is relaxed to
$d_i(X_i\setminus\{j\})\le\lambda \cdot d_i(X_h)$. Smaller values of
$\lambda$ give stronger fairness guarantees, with $\lambda=1$
corresponding to exact EFX.

The first non-trivial approximation for general additive chores was
obtained by \citet{zhou22efx}, who gave a polynomial-time algorithm
for computing an $O(n^2)$-EFX allocation. 
\citet{GMQ25} gave the first constant-factor existence result showing $4$-EFX allocations always exist. Their
approach was to define and prove the existence of fractional
\textit{earning-restricted (ER) equilibria}, round an ER equilibrium to a suitable
integral allocation, and perform chore swaps until obtaining a
$4$-EFX allocation. More recently, \citet{GM26} improved the existence
guarantee to $2$-EFX through a general framework for chore swaps.
Their framework begins with an allocation satisfying certain
properties and converts it to an approximate-EFX allocation in
polynomial time.

Despite these existence results, the question of computing a constant-approximate EFX allocation remained open. 
% As \citet[Section~5]{GM26} write, ``Another interesting question is if constant-EFX allocations can be computed in polynomial time.'' 
The main
difficulty in extending previous approaches lies in finding a suitable initial solution.
The approach of \citet{GMQ25} begins with an ER equilibrium, whose
existence proof uses a linear complementarity problem and a finite
pivoting procedure. The $2$-EFX result of \citet{GM26} instead uses
price-supported allocations that are envy-free up to one chore (EF1),
supplied by the existence theorem of
\citet{mahara}, whose proof uses a fixed-point argument. For a
constant number of agents, ER equilibria can be computed in polynomial
time~\citep{GMQ25}, as can EF1 and Pareto-optimal (PO)
allocations~\citep{mahara}. These initial allocations enable polynomial time computation of constant-EFX allocations when $n$ is constant.

For general $n$, however, these results do not give polynomial-time
algorithms for the starting solutions. In particular,
\citet[Section~2.4]{GMQ25}
identify approximate ER computation as the missing step that would
turn their existence results into polynomial-time algorithms, with
a small loss in the guarantees. Thus the constant-factor existence
results have left a substantial gap from the $O(n^2)$ approximation
attainable in polynomial time.

In this paper, we resolve this computational question for
additive chores and any number of agents.
\begin{theorem}\label{thm:main}
For any chore division instance with positive additive disutilities
and any $\eps\in(0,1)$, a $(3+\eps)$-EFX allocation
can be computed in time polynomial in the input length and $1/\eps$.
\end{theorem}

Our result relies on two algorithmic contributions. First, we
give a fully polynomial-time approximation scheme (FPTAS) for ER
equilibria, resolving the computational obstacle identified by
\citet{GMQ25}. Second, we give a new polynomial-time algorithm that
uses an approximate ER equilibrium to obtain a $(3+\eps)$-EFX
allocation. Our construction overcomes a factor-four bottleneck in
the earlier method for reallocating chores, which persists even
when the starting equilibrium is exact. We describe these
contributions below.

\paragraph{Computing an Approximate ER Equilibrium.}
In an earning restricted market, agent $i$ aims to earn $e_i$ by performing chores
of least disutility per unit of payment. Thus, agents are only assigned
chores that minimize their disutility-to-price ratio, called the
\textit{minimum pain-per-buck} (MPB) condition. Each chore $j$ has an earning
limit $c_j$ on the total amount it can pay to the agents. At prices
$p$, it pays out $\min\{p_j,c_j\}$; when its price exceeds this
limit, it is only partially allocated. 
When earning caps $c_j$ are large, this reduces to the standard Fisher market for chores. 

\citet{GMQ25} prove that an ER equilibrium exists if and only
if $\sum_jc_j\ge\sum_i e_i$, using a linear complementarity
formulation and a pivoting algorithm that terminates in finite time. However, polynomial-time
computation of even an approximate ER equilibrium remained open.
We give an FPTAS for ER markets with equal earning requirements $e_i=1$ and arbitrary positive earning limits satisfying
$\sum_jc_j\ge n$.

\begin{theorem}\label{thm:er-intro}
For every $\delta\in(0,1)$, an approximate ER equilibrium with
unit earning requirements and arbitrary feasible positive chore
limits can be computed in time polynomial in the input length and
$1/\delta$. Each agent earns an amount in $[1-\delta,1+\delta]$,
and MPB optimality, capped chore payouts, and total earnings hold exactly.
\end{theorem}

Earning limits separate a chore's price from its total
payout. In ordinary chore equilibria, every chore is fully allocated,
so prices sum to the total earning requirement $n$. With earning
limits, a chore can be partially allocated and its payout is
$q_j=\min\{p_j,c_j\}$, so prices need no longer sum to $n$.
We therefore work with the chore payouts $q_j$, which still sum
to $n$ and satisfy the simple bounds $q_j\le c_j$.

These payout variables also appear in the spending-restricted
extension of Shmyrev's convex program for goods
\citep{Shmyrev09,ColeEtAl17}. The program uses money flows
$b_{ij}$, denoting the amount agent $i$ spends on good $j$, and
total spending $q_j=\sum_i b_{ij}$. It
maximizes $\sum_{i,j}b_{ij}\log v_{ij}-\sum_jq_j\log q_j$
over nonnegative money flows $b$ and total spending $q$, subject to
each agent spending one and the caps $q_j\le c_j$; here $v_{ij}$ is agent $i$'s
value for good $j$. Thus, the
objective combines a term linear in the money flows with an
entropy term depending only on total spending.

A natural adaptation of the goods program to chores replaces
values by disutilities and maximization by minimization. Here,
$b_{ij}$ denotes the money agent $i$ earns from chore $j$, and
$q_j=\sum_i b_{ij}$ is its total payout. However, minimizing this
objective over both money flows and payouts need not give an
equilibrium. Even with one agent and two identical chores with
unit earning limits, it concentrates the entire payout on one
chore, leaving the other unpaid.

Our key insight is to minimize over money flows for fixed
payouts, while maximizing over the payouts themselves. For fixed
$q$, minimizing $\sum_{i,j}b_{ij}\log d_{ij}$, subject to each
agent earning one and each chore $j$ paying out $q_j$, is a
minimum-cost flow problem. Its optimal dual supplies prices
supporting MPB assignments. We then maximize this minimum flow
cost minus the entropy term $\sum_jq_j\log q_j$ over feasible
payouts. The resulting program, \textit{ER-MaxMin}, is nonconvex,
but we prove that every local maximum yields an exact ER
equilibrium.

To obtain an efficient algorithm, we apply mirror ascent to
this program. Each iteration computes an optimal flow and its
supporting MPB prices. The mirror-ascent step has a simple market
interpretation: scale these prices so that their capped payouts sum
to $n$. Since the new payouts can differ from those prescribed to
the flow computation, we rescale the money received from each chore
to match its new payout. This preserves MPB assignments and
satisfies the earning restrictions exactly, but may change
individual earnings. We stop when every agent's earning is within
$\delta$ of one; otherwise, we repeat with the new payouts.
The main analytical challenge is to relate these earning errors
to progress in the objective and thereby bound the number of
iterations, even when some chore payouts are arbitrarily small.
% We show that an earning error above $\delta$ forces the objective
% to increase by at least $\delta^2/4$,
% while its total possible increase is bounded.
Our analysis in \cref{sec:er-fptas} establishes polynomial-time
convergence, yielding an FPTAS for ER equilibria.

\paragraph{Constructing an EFX-Friendly Allocation.}
To obtain a $(3+\eps)$-EFX allocation from the approximate
equilibrium, we use the chore-swap framework of \citet{GM26}.
An agent can exchange one chore, called her \textit{swap chore},
for a least-disutility bundle while retaining her other chores,
which form her \textit{residual bundle}. The framework computes a
$\lambda$-EFX allocation in polynomial time from a suitable
starting allocation, called $\lambda$-\textit{EFX-friendly}.
In such an allocation, the \textit{inactive} agents are already
$\lambda$-envy-free, while each \textit{active} agent has one swap
chore and a residual bundle of small disutility relative to every
inactive bundle and every swap chore. Our task is therefore to
construct a $(3+\eps)$-EFX-friendly allocation.

\citet{GMQ25} show that an exact ER equilibrium can be
rounded to an integral allocation, which is subsequently modified
to yield a $4$-EFX-friendly allocation. There are two challenges in
extending this approach. First, our FPTAS gives only an approximate
ER equilibrium. Second, the reallocation must improve upon factor
four. We address the first challenge by extending the analysis of
their rounding algorithm to approximate earnings. With earning
limits $1/2$ and exact earnings, each agent receives at most two
chores of price above $1/2$. However, an arbitrarily small earning
error can cause a discrete jump from two such chores to three.

The earlier reallocation uses a fixed price threshold of
$1/2$ to determine which chores are eligible to serve as swap
chores. Unfortunately, this approach has a concrete bottleneck at
factor four, even when the starting equilibrium is exact. Our key
insight is to choose the swap chores and their recipients together
using a \textit{flexible} price threshold determined by the rounded
allocation. As the threshold increases, some chores initially
eligible for swaps rejoin their original residual bundles,
allowing agents with large residual bundles to remain inactive.
The challenge is to satisfy two requirements simultaneously:
every remaining swap chore needs a recipient with a sufficiently
small residual bundle, and every agent whose residual bundle has
price below the threshold must receive a swap chore. Our algorithm
ensures that every remaining swap chore has an eligible recipient;
a counting argument guarantees that enough chores remain for all
agents who need one.

Together with the rounding bounds, this construction gives
a $3$-EFX-friendly allocation from an exact ER equilibrium, with
the factor increasing by only $O(\delta)$ for a $\delta$-approximate
ER equilibrium. Choosing $\delta$ sufficiently small relative to
$\eps$ and applying the chore-swap framework yields a
$(3+\eps)$-EFX allocation in polynomial time, proving
\cref{thm:main}. We present the rounding analysis and the new
construction in \cref{sec:three-efx}.

\subsection{Related Work}\label{sec:related-work}

\paragraph{\normalfont \textit{EFX for restricted chore instances}.}
Several works obtain stronger EFX guarantees by restricting the number
of agents, the number of chores, or the disutility functions. For
$m\le2n$, \citet{KMS23} give a matching-based polynomial-time
algorithm for exact EFX; \citet{GMQ25} subsequently give a simpler
algorithm based on chore swaps. For three agents, \citet{AADM24}
compute a $2$-EFX allocation in polynomial time, even for subadditive
cost functions. For bivalued instances, where every disutility is in
$\{1,k\}$ for some $k>1$, \citet{LWZ25} compute a
$(2-1/k)$-EFX allocation while preserving fractional Pareto optimality.

\paragraph{\normalfont \textit{Fair and efficient chore allocations.}}
Prior to the general existence result of \citet{mahara}, polynomial-time
algorithms for simultaneously achieving EF1 and PO were known for
structured instances, including bivalued preferences
\citep{GMQ22,EPS22}. Mahara proves the existence of EF1 and fractionally
Pareto-optimal allocations for general additive chores, combining the
KKM lemma with a discrete allocation procedure. The fixed-point
argument supplies suitable supporting weights and prices, after which
the discrete procedure obtains the fairness guarantee. He also gives
a polynomial-time algorithm for EF1 and PO when the number of agents
is constant. His existence result supplies the starting point used by
\citet{GM26} to obtain $2$-EFX. Exact EFX and PO, however, can be
incompatible even for bivalued chores~\citep{HeTao2026}. Computing
EF1 and PO allocations in polynomial time for general instances
remains an interesting open question.

\paragraph{\normalfont \textit{Computing fractional chore equilibria.}}
For ordinary chore markets, \citet{BCM22} give an FPTAS for
competitive equilibrium with equal incomes using an exterior-point
method. \citet{CGMM22} give a combinatorial FPTAS based on price
updates and reallocations along MPB edges. \citet{CKMN24} develop a
dual equilibrium formulation and a greedy Frank--Wolfe algorithm whose
iterations solve linear programs. \citet{CJS24} give a log-price
difference-of-convex formulation, study a regularized algorithm, and
develop a smoothed gradient method. These works establish several
approaches for approximating chore equilibria, but their market model
requires full allocation of every chore. Earning restrictions instead
make the allocated supply of a capped chore depend on its price.
Our algorithm uses minimum-cost flows and a mirror-ascent step to
maintain exact MPB optimality and exact capped payouts, while allowing
small errors in individual earnings. \Cref{sec:er-flow} explains why
earning restrictions require a different formulation, and
\cref{sec:er-fptas} gives the algorithm and its analysis.

\paragraph{\normalfont \textit{Spending-restricted equilibria for goods.}}
The goods analogue of an earning limit is a spending limit on each
good. \citet{CG15} introduce spending-restricted equilibria and use
them to obtain a constant-factor approximation for maximizing Nash
social welfare with indivisible goods. \citet{ColeEtAl17} give convex
programming formulations of these markets, including an extension of
Shmyrev's money formulation~\citep{Shmyrev09}. Our max-min program
retains the capped payout domain and entropy term of this formulation.
The difference lies in agent optimality: for chores, the inner flow
problem minimizes the disutility-based cost, and its value is convex
in the payouts. The resulting outer maximization is nonconcave. We
use an optimal flow dual and a mirror-ascent step to increase its
objective, with progress controlled by the error in agent earnings.

\paragraph{\normalfont \textit{Independent work.}}
In independent contemporaneous work, \citet{QFLLC26} obtain the same
polynomial-time $(3+\eps)$-EFX guarantee. Their ``capped-price
structure'' reformulates the ER conditions of \citet{GMQ25} using
normalized chore shares and an approximate MPB condition. Their FPTAS uses multiplicative weight updates guided by max-flow-min-cut,
and is stated for a common earning limit for the chores. Our conceptual
contribution is a non-convex program whose local maxima
yield ER equilibria. Mirror ascent on this program gives an FPTAS, and moreover
maintains exact MPB and supports arbitrary earning limits. 

For the conversion from approximate ER to approximate EFX, their
``adaptive packet reallocation'' extends the chore-swap framework
of \citet{GM26} to agents with two swap chores. Our conversion
retains the rounding algorithm of \citet{GMQ25} and uses a flexible
price threshold to select swap chores and their recipients,
constructing an EFX-friendly allocation with one swap chore per
active agent.

\paragraph{Organization of the Paper.}
\Cref{sec:prelim} introduces our notation and formal definitions.
We describe and analyze the FPTAS for ER with equal earnings in \cref{sec:er-fptas}.
\Cref{sec:three-efx} gives the rounding analysis and the construction
of an EFX-friendly allocation. 
% The implementation details for the
% ER FPTAS are deferred to \cref{sec:er-bits}.

\section{Preliminaries}\label{sec:prelim}

Let $[n]=\{1,2,\ldots,n\}$ for a positive integer $n$.
\paragraph{Problem instance.}
A chore division instance $(N,M,D)$ consists of a set $N=[n]$ of
agents, a set $M=[m]$ of indivisible chores, and a list
$D=\{d_i\}_{i\in N}$ of additive disutility functions. We let
$d_{ij} \in \mathbb{Q}_{>0}$ denote the disutility of chore $j$ for agent $i$, and define the disutility ratio
$
 R=\max_{i\in N}\frac{\max_{j\in M}d_{ij}}{\min_{j\in M}d_{ij}}.$
For
$S\subseteq M$, we have $d_i(S)=\sum_{j\in S}d_{ij}$. 

An allocation
$X=(X_1,\ldots,X_n)$ is a partition of the chores, where $X_i$ denotes
the bundle assigned to agent $i$. A fractional allocation
$x\in\R_{\ge0}^{n\times m}$ assigns a fraction $x_{ij}$ of chore $j$
to agent $i$, with $\sum_i x_{ij}\le1$ for each chore $j$.
Agent $i$'s disutility under fractional allocation $x$ is $d_i(x_i)=\sum_jd_{ij}x_{ij}$.
Allocations are integral unless stated otherwise.

\paragraph{Fairness notions.}
For $\lambda\ge1$, agent $i$ is $\lambda$-envy-free ($\lambda$-EF) in allocation $X$ if $d_i(X_i)\le\lambda \cdot d_i(X_h)$ for every $h\in N$. Agent $i$ is $\lambda$-EFX in allocation $X$ if
\[
 d_i(X_i\setminus\{j\})\le\lambda \cdot d_i(X_h)
 \qquad\text{for every }h\in N\text{ and every }j\in X_i.
\]
An allocation is $\lambda$-EF or $\lambda$-EFX if the above condition holds for every agent. By definition, an agent assigned a singleton bundle is EFX. For a positive integer $k$, an allocation $X$ is $\lambda$-EF$k$ if, for every $i,h\in N$, there exists $S\subseteq X_i$ with $|S|\le k$ such that $d_i(X_i\setminus S)\le\lambda\cdot d_i(X_h)$. When $\lambda=k=1$, this becomes the well-studied EF1 fairness notion.

\paragraph{Prices and MPB allocations.}
Let $p=(p_1,\ldots,p_m)\in\R_{>0}^m$ be chore prices, where $p_j$ is the payment per unit of chore $j$. Let $p(S)=\sum_{j\in S}p_j$ and $p(x_i)=\sum_jp_jx_{ij}$ denote the prices of bundles $S$ and $x_i$, respectively. The \textit{minimum pain-per-buck} (MPB) ratio of agent $i$ and her set of MPB chores are
\[
 \alpha_i=\min_{j\in M}\frac{d_{ij}}{p_j},
 \qquad
 \mpb_i=\left\{j\in M:\frac{d_{ij}}{p_j}=\alpha_i\right\}.
\]
An allocation-price pair $(x,p)$ is MPB if
$x_{ij}>0$ implies $j\in\mpb_i$ for all $i,j$, i.e., an agent only performs chores that minimize her disutility to payment ratio. For an integral allocation,
this means $X_i\subseteq\mpb_i$ for every $i$.

Scaling an agent's disutilities by a positive constant preserves all her multiplicative fairness comparisons and her MPB chores. Thus, given an MPB allocation, we may divide agent $i$'s disutilities by
$\alpha_i$ and obtain
\begin{equation}\label{eq:mpb-normalization}
 d_{ij}\ge p_j\quad\text{for all }i,j,
 \qquad d_{ij}=p_j\quad\text{when } j\in\mpb_i.
\end{equation}
We use this normalization when analyzing the rounding and
chore-swap algorithms.

For an integral allocation $X$ and a positive integer $k$, let
\[
 p_{-k}(X_i)=\min_{S\subseteq X_i:\,|S|\le k}p(X_i\setminus S).
\]
This is the price remaining after removing up to $k$ highest-price
chores from $X_i$.
The pair $(X,p)$ is $\lambda$-price-EF$k$ ($\lambda$-pEF$k$) if
$p_{-k}(X_i)\le\lambda p(X_h)$ for all $i,h$. An MPB allocation
that is $\lambda$-pEF$k$ is also $\lambda$-EF$k$, by
\eqref{eq:mpb-normalization}. For $\lambda=k=1$, this becomes the price-EF1 (pEF1) condition introduced by \cite{Barman18FFEA} and used in several works on fair and efficient allocation \cite{barman2019prop1po,EPS22,GM23TCS,GMQ22,GMQ23chores,GMQ24wef1po,GMQ25,mahara2024polynomialtimealgorithmfairefficient,mahara}.

\paragraph{Pareto Optimality.}

An allocation $X$ is \textit{Pareto-optimal} (PO) if no integral
allocation $Y$ satisfies $d_i(Y_i)\le d_i(X_i)$ for every agent $i$,
with strict inequality for at least one agent. It is
\textit{fractionally Pareto-optimal} (fPO) if no complete fractional
allocation $y$ satisfies $d_i(y_i)\le d_i(X_i)$ for every agent $i$,
with strict inequality for at least one agent. Here, completeness means
that $\sum_i y_{ij}=1$ for every chore $j$. Thus fPO implies PO.

By the First Welfare Theorem \citep{mas1995microeconomic}, every
complete MPB allocation is fPO.
% To see this directly, under \eqref{eq:mpb-normalization}, the sum of disutilities is at least $p(M)$ for every complete fractional allocation and equals $p(M)$ at the MPB allocation. A Pareto improvement would strictly decrease this sum, which is impossible.
The Second Welfare Theorem \citep{mas1995microeconomic} gives the
converse: for every fPO allocation $X$, there exists a positive price vector $p$ such that $(X,p)$ is MPB. Thus an allocation is fPO if and only if it admits supporting MPB prices.

\subsection{Earning-Restricted Equilibria}

An \textit{earning-restricted} (ER) market $(N,M,D,e,c)$ consists of
a chore division instance $(N,M,D)$, agent earning requirements
$e_i\in\mathbb{Q}_{>0}$, and chore earning limits or caps
$c_j\in\mathbb{Q}_{>0}$.

Each agent $i$ aims to earn $e_i$ by performing chores of least
possible disutility, while the agents are restricted to collectively
receive at most $c_j$ from chore $j$. An equilibrium specifies a
fractional allocation and prices that meet these requirements.

\begin{definition}[ER equilibrium, \citealp{GMQ25}]\label{def:er}
A pair $(x,p)$ is an earning-restricted equilibrium if
\begin{enumerate}[label=(\roman*)]
\item $(x,p)$ is MPB and $p(x_i)=e_i$ for every agent $i$;
\item $\sum_i x_{ij}=\min\{1,c_j/p_j\}$ for every chore $j$.
\end{enumerate}
\end{definition}

Condition (i) requires each agent to earn her requirement using only MPB chores. It enforces agent optimality: the bundle $x_i$ minimizes $i$'s disutility among all bundles $z\ge 0$ that meet its earning requirement through $p(z)\ge e_i$.

Condition (ii) imposes the earning restriction: if $p_j\le c_j$, chore $j$ is fully allocated and pays out $p_j$;
otherwise, only a $c_j/p_j$ fraction is allocated and the payout is
$c_j$. \citet{GMQ25} show that an ER equilibrium exists if and only
if $\sum_i e_i\le\sum_jc_j$. Our algorithmic result concerns equal
earning requirements. Hence, we normalize them to $e_i=1$ for all $i$,
and assume $\sum_jc_j\ge n$.

We use the following approximation of ER equilibrium, which allows small errors in the agent earnings while retaining the MPB condition and the earning restrictions exactly.
\begin{definition}[Approximate ER equilibrium]\label{def:approx-er}
For $\delta\in(0,1)$, a pair $(x,p)$ is a $\delta$-approximate ER
equilibrium with unit earning requirements if
\begin{align}
 1-\delta\le p(x_i)\le1+\delta
 &\qquad(\forall i\in N),\label{eq:approx-earnings}\\
 x_{ij}>0\ \Longrightarrow\ j\in\mpb_i,
 &\qquad(\forall i\in N,\ \forall j\in M),\label{eq:approx-mpb}\\
 \sum_i x_{ij}=\min\{1,c_j/p_j\}
 &\qquad(\forall j\in M),\label{eq:approx-caps}\\
 \sum_i p(x_i)=n.&\label{eq:approx-total}
\end{align}
\end{definition}
\begin{remark}[The total earning constraint]
Summing \cref{eq:approx-earnings} over all agents gives $n(1-\delta)\le \sum_i p(x_i)\le n(1+\delta)$.
The usual relaxation of ER equilibrium allows the total earning to vary within this range, whereas \cref{eq:approx-total} requires it to equal $n$. Our algorithm maintains this stronger property, so we include it in the definition for convenience, and use it throughout the paper.
\end{remark}
% We will use this approximation notion throughout the paper.

% \paragraph{Agent Optimality.}
% For fixed prices $p$, the disutility-minimization problem of an agent
% who aims to earn $e_i$ and its LP dual are
% \begin{equation}\label{eq:agent-dual}
%  \min_{z\ge0}\left\{\sum_jd_{ij}z_j:\sum_jp_jz_j\ge e_i\right\}
%  =\max_{\gamma\ge0}\left\{\gamma e_i:
%                        \gamma p_j\le d_{ij}\ (j\in M)\right\}.
% \end{equation}
% The optimal dual variable is $\gamma=\min_j d_{ij}/p_j=\alpha_i$.
% By complementary slackness, an optimal bundle has earning exactly
% $e_i$ and only contains chores $j$ with $d_{ij}/p_j=\alpha_i$.
% This explains the MPB condition in \cref{def:er}. In an approximate
% ER equilibrium, agent $i$'s bundle is still disutility-minimizing
% for its actual earning $p(x_i)$, which need not be exactly one.

\paragraph{Money Flows.}
It is convenient to describe our algorithm and its analysis in terms of \textit{money flow} variables instead of allocation variables. Let $b_{ij}$ denote the earning of agent $i$ from chore $j$, and let
$q_j$ denote the total payout of chore $j$. For an allocation $(x,p)$,
these quantities are given by
\begin{equation}\label{eq:money-variables}
 b_{ij}=p_jx_{ij},\qquad q_j=\sum_i b_{ij}.
\end{equation}
In these variables, the agent earning requirement becomes
$\sum_jb_{ij}=e_i$, and the earning restriction becomes
$q_j=\min\{p_j,c_j\}$. Since prices are positive, $b_{ij}>0$ if
and only if $x_{ij}>0$, so the MPB condition is preserved as well.
Consequently, the approximate ER conditions can equivalently be
written as
\begin{equation}\label{eq:money-approx}
\begin{aligned}
 1-\delta\le\sum_jb_{ij}\le1+\delta
 &\quad(i\in N),\\
 b_{ij}>0\ \Longrightarrow\ j\in\mpb_i
 &\quad(i\in N,\ j\in M),\\
 \sum_i b_{ij}=q_j=\min\{p_j,c_j\}
 &\quad(j\in M),\\
 \sum_jq_j=n.&
\end{aligned}
\end{equation}
We will describe our equilibrium algorithm using these money flows.
When $b$ and $p$ are specified, each chore's payout is $q_j=\sum_i b_{ij}$,
and we abbreviate the solution as $(p,b)$. Conversely, given money
flows and prices satisfying the above conditions, define
$x_{ij}=b_{ij}/p_j$. The total fraction of chore $j$ allocated is
$q_j/p_j=\min\{1,c_j/p_j\}\le1$, and the earnings and MPB
assignments are unchanged. Thus $(x,p)$ satisfies
\cref{def:approx-er}.

\section{An FPTAS for ER Equilibria}\label{sec:er-fptas}

In this section, we design \cref{alg:er} for computing an approximate
ER equilibrium and prove its polynomial running time. We assume that
$e_i=1$ for every agent, that all disutilities and earning limits are
positive, and that $\sum_jc_j\ge n$. Our algorithm returns money
flows and prices satisfying \eqref{eq:money-approx}, which are
equivalent to the conditions in \cref{def:approx-er}. 
Let $R=\max_i(\max_j d_{ij}/\min_j d_{ij})$ denote the largest ratio between an agent's maximum and minimum disutility.

\begin{theorem}\label{thm:er-fptas}
Let $(N,M,D,e,c)$ be an ER market with unit earning requirements
and $\sum_jc_j\ge n$. For any $\delta\in(0,1)$, \cref{alg:er}
computes a $\delta$-approximate ER equilibrium using at most
\[
 1+\left\lceil\frac{4n\log(mR)}{\delta^2}\right\rceil
\]
minimum-cost flow computations. Each iteration takes polynomial time,
so the algorithm is an FPTAS. If $\sum_jc_j=n$, one minimum-cost flow
computation gives an exact ER equilibrium.
\end{theorem}

\subsection{A Mathematical Program for ER Equilibria}\label{sec:er-flow}

A natural source of inspiration for computing an approximate ER equilibrium lies in algorithms for competitive equilibrium (CE) of chores without earning limits~\citep{BCM22,CKMN24,CJS24}. These algorithms build on adaptations of convex programming formulations for goods. To build intuition, we first describe these formulations and the difficulties that arise when adapting them to ER equilibria.

Consider a goods market in which every agent has budget one and
value $v_{ij}>0$ for good $j$.\footnote{We assume strict positivity to simplify
the exposition, but it is straightforward to accommodate zero values too.} 
Write $v_i(x_i)=\sum_jv_{ij}x_{ij}$. 
A CE maximizes the Nash product of
agent values, or equivalently their sum of logarithms, as captured
by the Eisenberg--Gale (EG) convex program
\citep{Eisenberg1959EGConvexProgram}. The program and its dual are
shown below; we omit an additive constant from the dual objective.
\begin{equation}\label{eq:eg-goods-overview}
\begin{array}{c@{\qquad}c}
 \text{(EG)} & \text{(EG-Dual)}\\[4pt]
 \begin{aligned}[t]
  \max_x\quad&\sum_i\log v_i(x_i)\\
  \text{subject to}\quad&\sum_ix_{ij}\le1\quad(j\in M),\\
  &x\ge0
 \end{aligned}
 &
 \begin{aligned}[t]
  \min_{p,\beta}\quad&\sum_jp_j-\sum_i\log\beta_i\\
  \text{subject to}\quad&p_j\ge\beta_i v_{ij}\quad(i,j),\\
  &p\ge0,\quad\beta>0.
 \end{aligned}
\end{array}
\end{equation}
The primal variables describe the allocation, while the dual variables $p_j$ are the prices of goods and $\beta_i$ is the reciprocal of agent $i$'s \textit{maximum bang-per-buck} (MBB) ratio. Their optimality conditions ensure that every agent spends her budget on MBB goods and every good is fully allocated.

For chores, the analogous primal program minimizes the sum of log disutilities, and is shown below as EG-Chores. \citet{bogomolnaia2017manna} show that the KKT points of EG-Chores with positive disutility for every agent correspond to CE. In particular, every local minimum in this positive-disutility domain yields a CE. At these points, the supply constraints are tight and all chores are fully allocated.

\begin{equation}\label{eq:dual-chores-overview}
\begin{array}{c@{\qquad}c}
 \text{(EG-Chores)} & \text{(EG-Chores-Dual)}\\[4pt]
 \begin{aligned}[t]
  \min_x\quad&\sum_i\log d_i(x_i)\\
  \text{subject to}\quad&\sum_ix_{ij}\ge1\quad(j\in M),\\
  &x\ge0
 \end{aligned}
 &
 \begin{aligned}[t]
  \max_{p,\beta}\quad&\sum_jp_j-\sum_i\log\beta_i\\
  \text{subject to}\quad&p_j\le\beta_i d_{ij}\quad(i,j),\\
  &p\ge0,\quad\beta>0.
 \end{aligned}
\end{array}
\end{equation}

\citet{CKMN24} introduce a corresponding dual-like program by reversing the direction of optimization and the price inequalities in EG-Dual, shown above as EG-Chores-Dual.
Here $\beta_i$ represents the reciprocal MPB ratio, and the name EG-Chores-Dual refers to the analogy with goods; it is not a Lagrangian dual of EG-Chores. 
\citet{CKMN24} establish the corresponding KKT characterization for
EG-Chores-Dual.

However, both chore programs have difficulties for iterative
optimization. The primal objective tends to $-\infty$ as an agent's
disutility approaches zero, and the dual-like objective tends to
$+\infty$ as some $\beta_i$ approaches zero. These divergent values
near zero are called \emph{poles}. The dual-like objective is also
unbounded along directions in which prices increase. Thus improving
either objective can lead away from an equilibrium. To avoid these
directions, \citet{CKMN24} add the constraint $ \sum_jp_j=n$
% \begin{equation}\label{eq:ce-price-sum}
%  \sum_jp_j=n
% \end{equation}
to EG-Chores-Dual. This constraint is redundant at CE since all chores are fully allocated and each agent earns one. \citet{CKMN24} show that adding it preserves the KKT characterization while removing the poles and unbounded improving directions, thereby enabling an algorithm for approximate CE.

Unfortunately, it is not clear how to extend these formulations to accommodate earning restrictions. In an ER equilibrium, only a $\min\{1,c_j/p_j\}$ fraction of chore $j$ is allocated, so the total earning condition becomes $\sum_j\min\{p_j,c_j\}=n$. Adding $p_j\le c_j$ would bound prices rather than payouts and could exclude valid ER equilibria with $p_j>c_j$. Moreover, the price-sum constraint $\sum_j p_j = n$ may exclude every ER equilibrium, as the following example shows.

\begin{example}[Price normalization]\label{ex:normalization}
Consider two chores and one agent with disutilities $(1,4)$, earning requirement one,
and earning limits $(1/2,1/2)$. The ordinary chore equilibrium has
prices $(1/5,4/5)$. If we impose the earning limits at these prices,
the total payout is $1/5+1/2=7/10$, and the agent cannot earn one.
In an ER equilibrium, both chores must pay out $1/2$, and the MPB
condition requires their prices to be in ratio $1:4$. Thus prices
$(1/2,2)$ support an ER equilibrium with allocation $(1,1/4)$.
Moreover, every ER price vector has price sum at least $5/2$. Hence,
imposing a price sum of one would exclude every ER equilibrium.
\end{example}

Earning limits also prevent a direct adaptation of price updates that preserve the allocated quantities by scaling prices and money flows together. Once a chore's payout reaches its earning limit, increasing its price must reduce its total quantity allocated.
\begin{samepage}
\begin{example}[Scaling prices and money flows]\label{ex:scaling}
Consider two chores and two agents with identical disutilities $(1,3)$, unit earning requirements, and earning limits $(1,1)$. Prices $(1,3)$ and earnings $b_{ij}=1/2$ for all $i,j$ form an ER equilibrium. The first chore is fully allocated, while one-third of the second chore is allocated. Doubling the second chore's price and all earnings from it preserves this allocation, but increases its payout to two and violates its earning limit. If we instead keep its payout at one, doubling its price halves the quantity allocated.
\end{example}
\end{samepage}

The two examples suggest that we should work with the total payout
$q_j$ of a chore $j$, keeping it distinct from its price $p_j$. In these
variables, we can impose the earning limits directly through
$q_j\le c_j$ and maintain the total payout $\sum_jq_j=n$. We must
then find supporting MPB prices that satisfy
$q_j=\min\{p_j,c_j\}$.

\paragraph{Connection to Spending-Restricted Equilibria for Goods.}
For goods, the \textit{spending-restricted} (SR) equilibrium has precisely this relation between prices and total spending. In this model, an agent with unit budget buys goods to maximize her value, while the total spending on good $j$ is restricted to $c_j$. If its price exceeds $c_j$, only a $c_j/p_j$ fraction of the good is sold. \citet{ColeEtAl17} connect this model to the convex program of \citet{Shmyrev09}. Applying the change of variables $(p,\beta)\mapsto(\log p,-\log\beta)$ to EG-Dual and then taking its dual gives Shmyrev's formulation for ordinary goods markets. Its variables are $q_j$ and $b_{ij}$, denoting the total spending on good $j$ and the money agent $i$ spends on good $j$, respectively. Since this program has money-flow variables, it can be adapted to incorporate spending limits.
% The exponential price terms give conjugates $q_j\log q_j-q_j$, explaining the entropy term in this formulation; the linear terms sum to the constant $n$.
Adding the constraints
$q_j\le c_j$ to Shmyrev's program gives the SR program of
\citet{ColeEtAl17}:
\begin{equation}\label{eq:shmyrev-overview}
\begin{aligned}
 \max_{b,q}\quad&
       \sum_{i,j}b_{ij}\log v_{ij}-\sum_jq_j\log q_j\\
 \text{subject to}\quad&\sum_jb_{ij}=1 &&(i\in N),\\
 &\sum_ib_{ij}=q_j                   &&(j\in M),\\
 &0\le q_j\le c_j                    &&(j\in M),\\
 &b_{ij}\ge0                         &&(i\in N,\ j\in M).
\end{aligned}
\end{equation}
We use the convention $0\log0=0$ for the entropy term $\mathcal H(q) = \sum_j q_j \log q_j$. The first two constraints already imply $\sum_jq_j=n$. 
% Removing the upper bounds $q_j\le c_j$ gives the ordinary Shmyrev program, whose optimal spending totals are the equilibrium prices. 
With these bounds, prices are recovered from the dual variables and satisfy $q_j=\min\{p_j,c_j\}$. The optimality conditions also ensure that agents buy only MBB goods.

\paragraph{Adapting Shmyrev's Program to Chores.}
Similar to EG-Chores-Dual, an adaptation of the SR program to chores replaces values by disutilities and maximization by minimization, giving
\begin{equation}\label{eq:shmyrev-chores}
\begin{gathered}
\text{(Shmyrev-Chores)}\\[4pt]
\begin{aligned}
 \min_{b,q}\quad&
       \sum_{i,j}b_{ij}\log d_{ij}-\sum_jq_j\log q_j\\
 \text{subject to}\quad&\sum_jb_{ij}=1 &&(i\in N),\\
 &\sum_ib_{ij}=q_j                   &&(j\in M),\\
 &0\le q_j\le c_j                    &&(j\in M),\\
 &b_{ij}\ge0                         &&(i\in N,\ j\in M).
\end{aligned}
\end{gathered}
\end{equation}
At first glance, this program may seem promising since it has a bounded objective, unlike EG-Chores. However, its optima need not be ER equilibria! 

\begin{example}
Consider two chores and one agent with
disutilities $(1,1)$ and earning limits $(1,1)$. Its objective is
$-q_1\log q_1-q_2\log q_2$, which is minimized at payouts $(1,0)$
or $(0,1)$. However, the ER equilibrium has payouts and prices
$(1/2,1/2)$. 
\end{example}

Thus reversing the direction of optimization for the entire program does not give the desired characterization. This example isolates the choice of chore payouts: with only one agent, the money she receives from each chore is necessarily $b_{1j}=q_j$. Minimizing the objective concentrates all the payout on one chore, leaving the other with no payment. 
On the other hand, maximizing it gives equal payouts, as required for these identical chores.
The reason for maximizing becomes clearer when the chores have different disutilities. A payout specifies how much money a chore provides; it is not the disutility incurred by an agent. An uncapped chore with higher disutility must have a proportionally higher price to be equally attractive to the agent. The following example shows how the maximization achieves this and how earning limits change the resulting allocation.

\begin{example}[Choosing chore payouts]\label{ex:payout-choice}
Consider two chores and one agent with disutilities $(1,2)$ and earning limits
$(1,1)$. Since $b_{1j}=q_j$, maximizing the objective amounts to
\[
\begin{aligned}
 \max_{q_1,q_2}\quad&
 q_1\log\frac{1}{q_1}+q_2\log\frac{2}{q_2}\\
 \text{subject to}\quad&q_1+q_2=1,\\
 &q_1,q_2\ge0.
\end{aligned}
\]
The maximizer is $(q_1,q_2)=(1/3,2/3)$. Taking these payouts as
prices makes both chores MPB, since
$d_{11}/p_1=d_{12}/p_2=3$. Both chores are fully allocated and
the agent earns one.

Now reduce the second chore's earning limit to $1/2$. Maximizing
the same objective subject to this cap gives payouts $(1/2,1/2)$.
These payouts cannot themselves serve as prices, since at equal
prices the agent would prefer the first chore. Instead, prices
$(1/2,1)$ make both chores MPB and induce precisely these payouts.
The agent performs all of the first chore and half of the second.
Thus the capped chore retains a higher price even though its payout
cannot exceed $1/2$.
\end{example}

The above examples suggest that maximizing over chore payouts, rather than minimizing, is the better choice.
Unfortunately, the picture is more complicated with several agents, since we must also decide how to distribute each chore's payout among them. Agent $i$ receives $b_{ij}$ of the total payout $q_j$ from chore $j$, and earns one in total. For fixed payouts $q$, we must choose the money flows $b_{ij}$ to ensure agent optimality, i.e., the MPB condition. We show that minimizing $\sum_{i,j}b_{ij}\log d_{ij}$ over these money flows achieves this purpose: its optimal flows are exactly those that admit supporting MPB prices. Having obtained these assignments, we choose the chore payouts by maximizing the resulting value minus the entropy term, as motivated by \cref{ex:payout-choice}.

\paragraph{A Mathematical Program for ER Equilibrium.}
With the above intuition, we introduce our program for ER equilibria. We set up some notation first. Let $Q$ be the set of feasible chore payouts:
\begin{equation}\label{eq:payout-domain}
 Q=\Big\{q\in\R_{\ge0}^m:q_j\le c_j,\ \sum_jq_j=n\Big\}.
\end{equation}
For fixed $q\in Q$, we distribute the payouts by solving the following program:
\begin{equation}\label{eq:flow-value}
\begin{aligned}
 C(q)=\min_{b\ge0}\quad&\sum_{i,j}b_{ij}\log d_{ij}\\
 \text{subject to}\quad&\sum_jb_{ij}=1 &&(i\in N),\\
 &\sum_ib_{ij}=q_j &&(j\in M).
\end{aligned}
\end{equation}
This is a minimum-cost flow problem: the constraints require each agent to earn one and each chore $j$ to pay out $q_j$, and sending one unit from chore $j$ to agent $i$ has cost $\log d_{ij}$. The problem is feasible for every $q\in Q$, since $b_{ij}=q_j/n$ satisfies these constraints. Let $\mathcal B(q)$ denote its set of feasible money flows.

Our program for ER equilibria, termed \textit{ER-MaxMin}, is presented below.
\begin{equation}\label{eq:potential}
\begin{gathered}
\text{(ER-MaxMin)}\\[4pt]
\begin{aligned}
 \max_q\quad&
 \left\{\min_{b\in\mathcal B(q)}\sum_{i,j}b_{ij}\log d_{ij}
                            -\sum_jq_j\log q_j\right\}\\
 \text{subject to}\quad&\sum_jq_j=n,\\
 &0\le q_j\le c_j &&(j\in M).
\end{aligned}
\end{gathered}
\end{equation}

Unlike the SR program for goods \cref{eq:shmyrev-overview}, ER-MaxMin is generally nonconvex. Nevertheless, we prove that \textit{every local maximum of the ER-MaxMin program can be transformed to an ER equilibrium}. To prove this, we first establish the MPB guarantee of the solution to the flow problem \cref{eq:flow-value} via its dual. Writing its variables as $\log\alpha_i$ and $\log w_j$ corresponding to the agent and chore constraints in \cref{eq:flow-value}, we obtain
\begin{equation}\label{eq:flow-dual}
\begin{aligned}
 C(q)=\max_{\alpha,w}\quad&\sum_i\log\alpha_i+\sum_jq_j\log w_j\\
 \text{subject to}\quad&\alpha_iw_j\le d_{ij}
                                      &&(i\in N,\ j\in M),\\
 &\alpha>0,\quad w>0.
\end{aligned}
\end{equation}
This is a linear program in $\log\alpha_i$ and $\log w_j$.
The logarithmic costs are what give the desired MPB guarantee: the additive dual inequalities become $\alpha_iw_j\le d_{ij}$ after exponentiation.

\begin{lemma}[Minimum-cost flows and MPB prices]\label{lem:flow-duality}
For every $q\in Q$, a flow $b\in\mathcal B(q)$ is optimal if and only if it admits supporting MPB prices. Every optimal dual $(\alpha,w)$ supplies such prices $w$ for every optimal flow, with $\alpha_i$ equal to agent $i$'s MPB ratio.
\end{lemma}
\begin{proof}
For any optimal flow $b$ and optimal dual $(\alpha,w)$,
complementary slackness gives
\[
 \frac{d_{ij}}{w_j}\ge\alpha_i,\qquad
 b_{ij}>0\ \Longrightarrow\ \frac{d_{ij}}{w_j}=\alpha_i.
\]
Thus agents receive money only from their MPB chores. Conversely,
if a feasible flow $b\in\mathcal B(q)$ is MPB at some prices $w$,
then its MPB ratios give a feasible dual that is tight on every
positive flow edge. Hence $b$ is a minimum-cost flow. The inner
minimization therefore captures precisely the assignments that
admit supporting MPB prices.
\end{proof}

\Cref{lem:flow-duality} shows that the inner minimization enforces MPB optimality. It remains to choose payouts that are consistent with their supporting prices. The constraint $q_j\le c_j$ only bounds a chore's payment; an ER equilibrium additionally requires $q_j=\min\{p_j,c_j\}$. We show that local optimality of the outer maximization enforces this condition. For this, we first establish additional properties of the ER-MaxMin program. Let $V(q)=C(q)-\mathcal H(q)$ denote its objective, where $\mathcal{H}(q) = \sm_j q_j\log q_j$. 

For fixed positive prices $w$, define $\Phi_w(z)=\sum_jz_j\log w_j-\mathcal H(z)$ over $z\in Q$.
% \[
%  \Phi_w(z)=\sum_jz_j\log w_j-\mathcal H(z)\qquad(z\in Q).
% \]

\begin{lemma}[Convexity and a concave lower bound]\label{lem:flow-lower-bound}
The function $C$ is convex on $Q$. For any $q\in Q$,
any optimal dual $(\alpha,w)$ at $q$, and every $z\in Q$,
\begin{equation}\label{eq:flow-support}
 C(z)\ge C(q)+\sum_j(z_j-q_j)\log w_j.
\end{equation}
Thus $\log w$ is a subgradient of $C$ at $q$. Moreover, the function
\begin{equation}\label{eq:flow-minorant}
\begin{aligned}
 F_q(z)&=C(q)+\sum_j(z_j-q_j)\log w_j-\mathcal H(z)\\
       &=V(q)+\Phi_w(z)-\Phi_w(q)\qquad(z\in Q).
\end{aligned}
\end{equation}
is concave and satisfies $F_q(z)\le V(z)$ for all $z\in Q$, with
$F_q(q)=V(q)$.
\end{lemma}
\begin{proof}
An optimal dual $(\alpha,w)$ at $q$ remains
feasible for every $z\in Q$, so
\[
 C(z)\ge\sum_i\log\alpha_i+\sum_jz_j\log w_j
      =C(q)+\sum_j(z_j-q_j)\log w_j.
\]
This proves \eqref{eq:flow-support} and the subgradient claim.
By \eqref{eq:flow-dual}, $C$ is the pointwise maximum of affine
functions of the payouts, so it is convex.
Finally, \eqref{eq:flow-support} gives $F_q(z)\le V(z)$, with
equality at $q$. The function $F_q$ is concave because it is an
affine function minus the convex function $\mathcal H$.
\end{proof}

% Since $C$ is convex, but $V=C-\mathcal H$ need not be concave. Thus ER-MaxMin is generally a nonconvex program.
% In the goods SR program, the corresponding maximum-cost flow value is $ G(q)=\max_{b\in\mathcal B(q)}\sum_{i,j}b_{ij}\log v_{ij}
% $. The function $G$ is concave, so maximizing $G-\mathcal H$ is a convex
% program. For chores, 

\begin{proposition}[Local maxima yield ER equilibria]\label{prop:er-fixed-point}
Every local maximizer $q$ of ER-MaxMin \eqref{eq:potential} is positive. For any optimal flow $b$ and dual $(\alpha,w)$ at $q$, there is a scalar $\tau>0$ such that $(\tau w,b)$ is an exact ER equilibrium and $q_j=\min\{\tau w_j,c_j\}$ for every chore $j$.
\end{proposition}
\begin{proof}
Fix a local maximizer $q$ and optimal flow and dual solutions $b$ and $(\alpha,w)$.
By \cref{lem:flow-lower-bound}, $F_q$ is a lower bound on $V$
that is tight at $q$. Hence $q$ locally maximizes $F_q$, and therefore
also $\Phi_w$, since the two functions differ by a constant.

We show that $q$ must be positive. If $q_j=0$, transfer an amount
$\varepsilon>0$ to coordinate $j$ from a positive coordinate.
Since $c_j>0$, this is feasible for sufficiently small
$\varepsilon$. The change in $\Phi_w$ is
$-\varepsilon\log\varepsilon+O(\varepsilon)>0$, contradicting
optimality.

The first-order optimality conditions for $\Phi_w$ on the polytope
$Q$ now give a scalar $\nu$ and multipliers $\mu_j$ satisfying
\[
 \log w_j-\log q_j-1=\nu+\mu_j,\qquad
 \mu_j\ge0,\qquad \mu_j(c_j-q_j)=0.
\]
Set $\tau=e^{-\nu-1}$ and $p_j=\tau w_j$. Then
$p_j=q_je^{\mu_j}$. If $q_j<c_j$, complementary slackness gives
$\mu_j=0$ and $p_j=q_j$; if $q_j=c_j$, we have $p_j\ge c_j$.
Therefore $q_j=\min\{p_j,c_j\}$ in both cases. By
\cref{lem:flow-duality}, the flow $b$ is MPB at $w$, and hence at
$p=\tau w$. Since $b$ also gives every agent earning one enforced by the constraint in \cref{eq:flow-value}, $(p,b)$
is an exact ER equilibrium.
\end{proof}

% Since $Q$ is compact and $C$ is continuous, $V$ attains a maximum.
% \Cref{prop:er-fixed-point} therefore also proves the existence of an ER equilibrium with equal earnings. 
Thus, finding a local maximum of ER-MaxMin suffices to compute an exact ER equilibrium. In the next subsection, we use \textit{mirror ascent} on ER-MaxMin to compute an \textit{approximate} ER equilibrium in polynomial time.

\subsection{Computing an Approximate ER Equilibrium}\label{sec:er-mirror}

Mirror ascent is an iterative first-order method for constrained
maximization. At each step, it maximizes a first-order approximation
of the objective, with a penalty for moving away from the current
point. The penalty is a Bregman divergence generated by a convex
function called the \emph{mirror map}~\citep{BeckTeboulle03}.
For ER-MaxMin, we use the entropy function $\mathcal H$ as the mirror map. 
As we show below, the resulting mirror-ascent step maximizes the concave lower bound $F_{q}$ from \cref{lem:flow-lower-bound}.

To define the update, fix a positive payout vector $q\in Q$ and solve the flow program \eqref{eq:flow-value} to obtain an optimal flow $b$ and supporting
prices $w$ from its dual. By \cref{lem:flow-lower-bound}, $\log w$ is a subgradient
of $C$ at $q$. Subtracting the gradient of $\mathcal H$ gives the
direction $g_j=\log w_j-\log q_j-1$ for the first-order
approximation of $V$. When $C$ is differentiable, $g=\nabla V(q)$;
otherwise, this construction uses the subgradient supplied by the
flow dual. The
divergence generated by $\mathcal H$ on $Q$ is the relative entropy
\begin{equation}\label{eq:er-kl}
\begin{aligned}
 \KL(z\Vert q)
 &=\sum_jz_j\log(z_j/q_j)\\
 &=\mathcal H(z)-\mathcal H(q)
      -\sum_j(\log q_j+1)(z_j-q_j)\qquad(z\in Q).
\end{aligned}
\end{equation}
The second equality uses $\sum_j(z_j-q_j)=0$. Rearranging this
identity for $\mathcal H(z)$ and substituting into the lower bound
from \cref{lem:flow-lower-bound} gives
\begin{equation}\label{eq:mirror-model}
\begin{aligned}
 V(z)\ge F_q(z)
 &=C(q)+\sum_j(z_j-q_j)\log w_j-\mathcal H(z)\\
 &=V(q)+\sum_j(\log w_j-\log q_j-1)(z_j-q_j)
          -\KL(z\Vert q)\\
 &=V(q)+\sum_jg_j(z_j-q_j)-\KL(z\Vert q).
\end{aligned}
\end{equation}
Thus, maximizing $F_q$ over $Q$ is a mirror-ascent step with step
size one. Let $r\in Q$ be the updated payouts computed by maximizing $F_q$ over $Q$. By \cref{eq:flow-minorant}, $F_q$ differs from $\Phi_w$ by a constant, so computing $r$ amounts to maximizing
$\Phi_w$ over $Q$.
The proof of \cref{prop:er-fixed-point} shows that, at a local
maximum of ER-MaxMin, the current payouts $q$ already maximize
$\Phi_w$. The mirror-ascent update performs this maximization at
each iteration. Therefore, the same optimality conditions give the
updated payouts $r$ by scaling the supporting prices and truncating
at the earning limits, as stated in the following lemma.

\begin{lemma}[Mirror-ascent step]\label{lem:entropy}
For positive numbers $w_j$, the problem
\begin{equation}\label{eq:entropy-update}
\begin{aligned}
 \max_{r\in Q}\quad&\Phi_w(r)=\sum_jr_j\log w_j-\mathcal{H}(r)
\end{aligned}
\end{equation}
has a unique maximizer. It is positive and has the form $r_j=\min\{c_j,\tau w_j\}$ for some $\tau>0$, so that $\sum_j\min\{c_j,\tau w_j\}=n$. For rational $w$, such a $\tau$ and vector $r$ can be computed in polynomial time.
\end{lemma}
\begin{proof}
As $\tau$ increases from zero, $\sum_j\min\{c_j,\tau w_j\}$
increases continuously from zero to $\sum_jc_j\ge n$, so a suitable
$\tau>0$ exists. The resulting $r$ is positive and satisfies the
optimality conditions used in \cref{prop:er-fixed-point}, with
$\nu=-\log\tau-1$ and $\mu_j=\log(\tau w_j/r_j)\ge0$.
Strict concavity of $\Phi_w$ proves uniqueness.

We show how to compute $\tau$. If  $\sum_jc_j=n$, taking $\tau=\max_jc_j/w_j$ makes every chore pay its cap, so $r=c$. Therefore, suppose $\sum_jc_j>n$. Sort and relabel the chores so that $c_1/w_1\le\cdots\le c_m/w_m$, and consider the intervals $\left[0,\frac{c_1}{w_1}\right),\quad \left[\frac{c_1}{w_1},\frac{c_2}{w_2}\right),\quad\ldots,\quad
 \left[\frac{c_m}{w_m},\infty\right)$. 
Index these intervals by $k=0,\ldots,m$. If $\tau$ lies in interval $k<m$, the first $k$ chores are capped and the remaining chores pay $\tau w_j$. Thus the normalization condition becomes the linear equation $ \sum_{j\le k}c_j+\tau\sum_{j>k}w_j=n,$
whose solution is
\begin{equation}\label{eq:scale-rational}
 \tau=\frac{n-\sum_{j\le k}c_j}{\sum_{j>k}w_j}.
\end{equation}
For each of the first $m$ intervals, compute this value and check
whether it belongs to the interval. The final interval cannot
contain a solution, since all chores are capped there and their
total payout is $\sum_jc_j>n$. Hence one of the checks must succeed in finding $\tau$, and we return $r_j=\min\{c_j,\tau w_j\}$. 
% Empty intervals caused by tied ratios can be skipped. 
Sorting and checking these intervals
take polynomial time for rational $w$ and $c$.
\end{proof}

The mirror-ascent step is useful since it can only improve the objective: the updated payouts $r$ satisfy $V(r)\ge F_q(r)\ge F_q(q)=V(q)$ due to \cref{lem:flow-lower-bound}.
      
Consider a mirror-ascent step at payouts $q$ and corresponding optimal money flows $b$. Let $r$ be the updated payouts and $p = \tau w$ be the prices computed using \cref{lem:entropy}.
If $r=q$, then $(p,b)$ is an exact ER equilibrium. In general, we match the new payouts by rescaling the money flow $b$ as
\begin{equation}\label{eq:er-rescaling}
 \widehat b_{ij}=b_{ij}\frac{r_j}{q_j},\qquad
 \widehat e_i=\sum_j\widehat b_{ij}.
\end{equation}
This preserves MPB assignments because the rescaling leaves
the support of $b$ unchanged, and uniformly scaling prices from $w$
to $p=\tau w$ preserves each agent's MPB chores. Moreover,
$\sum_i\widehat b_{ij}=(r_j/q_j)\sum_i b_{ij}=r_j=\min\{c_j,p_j\}$,
as required by the earning limits. Individual earnings, however,
may change from one to $\widehat e_i$.  
Thus we return $(p,\widehat b)$ if every $\widehat e_i$ lies in
$[1-\delta,1+\delta]$; otherwise, we set $q\gets r$ and begin
the next iteration.

To initialize the iteration, we apply \cref{lem:entropy} to uniform weights
$w_j=1$. This gives a positive payout vector, and every subsequent
vector is positive by the same lemma, so the rescaling in
\cref{eq:er-rescaling} is
well-defined. \Cref{alg:er} collects these steps.

\begin{algorithm}[t]
\caption{Approximate ER equilibrium with equal earnings}\label{alg:er}
\textbf{Input:} Instance $(N,M,D,e=1,c)$ with $\sum_jc_j\ge n$;
accuracy $\delta\in(0,1)$\\
\textbf{Output:} A $\delta$-approximate ER equilibrium $(p,b)$
\begin{algorithmic}[1]
\State Initialize $q_j\gets\min\{c_j,\tau_0\}$, choosing $\tau_0$ so $\sum_jq_j=n$
\Loop
\State Solve \eqref{eq:flow-value} at $q$, obtaining an optimal flow $b$ and supporting prices $w$
\State Compute the price scale $\tau$ using \cref{lem:entropy}
\State Set $p_j\gets\tau w_j$ and $r_j\gets\min\{c_j,p_j\}$ for every $j$
\State Set $\widehat b_{ij}\gets b_{ij}r_j/q_j$ and $\widehat e_i\gets\sum_j\widehat b_{ij}$
\If{$|\widehat e_i-1|\le\delta$ for every $i\in N$}
\State \Return $(p,\widehat b)$
\EndIf
\State $q\gets r$
\EndLoop
\end{algorithmic}
\end{algorithm}

\begin{lemma}\label{lem:er-correct}
Every candidate $(p,\widehat b)$ of \cref{alg:er} satisfies exact MPB
optimality, capped chore payouts, and total earning $n$. If the
stopping condition holds, it is a $\delta$-approximate ER equilibrium.
\end{lemma}
\begin{proof}
By \cref{lem:flow-duality}, $b$ is MPB at $w$, and hence
at $p=\tau w$. The rescaling in \cref{eq:er-rescaling} preserves its support, and
\[
 \sum_i\widehat b_{ij}
 =\frac{r_j}{q_j}\sum_ib_{ij}=r_j=\min\{c_j,p_j\}.
\]
These payouts sum to $n$. When the stopping condition on agent earning holds, all conditions defining an approximate ER equilibrium \eqref{eq:money-approx} are therefore satisfied.
\end{proof}

\subsection{Bounding the Number of Iterations}

We now bound the number of iterations of \cref{alg:er} by showing that each nonterminating iteration increases $V$ by a definite amount. 
% We then bound the total increase available to $V$.
Without loss of generality, we assume $1\le d_{ij}\le R$ for all $i\in N,\ j\in M$.

\begin{lemma}[Progress in each iteration]\label{lem:progress}
Every iteration of \cref{alg:er} that does not terminate increases
$V$ by at least $\delta^2/4$.
\end{lemma}
\begin{proof}
Let $q$ and $r$ be the payouts before and after a mirror-ascent step, and let $w$ be the supporting prices at $q$. By \cref{lem:entropy}, $r$ maximizes $\Phi_w$ over $Q$, so first-order optimality gives
$\langle\nabla\Phi_w(r),q-r\rangle\le0$.
Using \cref{lem:flow-lower-bound} and the entropy identity
\eqref{eq:er-kl} at $r$, we obtain
\begin{equation}\label{eq:progress}
\begin{aligned}
 V(r)-V(q)
 &\ge\Phi_w(r)-\Phi_w(q)\\
 &=\KL(q\Vert r)-\langle\nabla\Phi_w(r),q-r\rangle\\
 &\ge\KL(q\Vert r).
\end{aligned}
\end{equation}

We next relate this bound to the agent earnings after rescaling.
A small payout $q_j$ can make the ratio $r_j/q_j$ large, so we
apply the log-sum inequality to the payments each agent
receives from the chores. Let $b$ be the unit-earning flow and
$\widehat b_{ij}=b_{ij}r_j/q_j$ its rescaling, and define
$\omega(z)=z-1-\log z$ for $z>0$. Then
\begin{align*}
 \KL(q\Vert r)
 &=\sum_{i,j:b_{ij}>0}b_{ij}
       \log\frac{b_{ij}}{\widehat b_{ij}}\\
 &\ge\sum_i\bigg(\sum_jb_{ij}\bigg)
       \log\frac{\sum_jb_{ij}}{\sum_j\widehat b_{ij}}\\
 &=-\sum_i\log\widehat e_i
 =\sum_i\omega(\widehat e_i).
\end{align*}
The first equality follows from the rescaling and
$\sum_i b_{ij}=q_j$. The last two equalities use the original unit
earnings and $\sum_i\widehat e_i=n$, respectively.

The function $\omega$ is nonnegative and vanishes only at one.
For $\delta\in(0,1)$, if $|z-1|>\delta$, then
\begin{equation}\label{eq:omega}
 \omega(z)\ge\delta-\log(1+\delta)\ge\delta^2/4.
\end{equation}
Indeed, $\omega$ decreases below one and increases above one, and
$\delta-\log(1+\delta)=\int_0^\delta t/(1+t)\,dt\ge\delta^2/4$;
the corresponding deviation below one has at least this value.
Since the iteration does not terminate, some agent has
$|\widehat e_i-1|>\delta$. Hence
$\KL(q\Vert r)\ge\sum_i\omega(\widehat e_i)\ge\delta^2/4$,
and \eqref{eq:progress} proves the claim.
\end{proof}

\begin{proof}[\bf Completing the proof of \cref{thm:er-fptas}]
We first bound the total increase available to $V$.
Since $1\le d_{ij}\le R$, we have $0\le C(q)\le n\log R$.
Convexity of $z\log z$, together with $\sum_jq_j=n$ and $q_j\le n$,
gives $n\log(n/m)\le\mathcal H(q)\le n\log n$.
Since $V=C-\mathcal H$, it follows that
$-n\log n\le V(q)\le n\log R-n\log(n/m)$ for all $q\in Q$.
Thus the range of $V$ over $Q$ is at most $n\log(mR)$.

By \cref{lem:progress}, each nonterminating iteration increases $V$
by at least $\delta^2/4$. Hence there are at most
$4n\log(mR)/\delta^2$ such iterations. Including the final iteration
gives the bound in \cref{thm:er-fptas}. If $\sum_jc_j=n$, then
$Q=\{c\}$ and every update has $q=r=c$, so the first iteration
returns an exact ER equilibrium. Together with \cref{lem:er-correct},
this proves correctness and the polynomial iteration bound of \cref{alg:er}. 

However, we must address two implementation challenges before obtaining polynomial running time. First, the costs $\log d_{ij}$ may be
irrational, yet each iteration requires an exact optimal flow and
supporting MPB prices. Second, the encoding lengths of the maintained
flows, prices, and payouts must remain polynomially bounded throughout
the iterations. We show how to handle these issues in \cref{sec:er-bits} so that
each iteration takes polynomial time. Together with the iteration
bound, this completes the proof of \cref{thm:er-fptas}.
\end{proof}

\begin{remark}[Non-identical earning requirements]\label{rem:unequal}
We note that the same algorithm can be run with arbitrary requirements $e_i>0$,
replacing each agent's unit earning requirement by $e_i$ and total payout $n$ by $E=\sum_i e_i$. The earning-error argument in \cref{lem:progress} then guarantees progress
of at least $e_{\min}\delta^2/4$, where $e_{\min}=\min_i e_i$.
This yields $O(1+(E/e_{\min})\log(mR)/\delta^2)$ iterations.
The ratio $E/e_{\min}$ may be exponential in the binary input length.
Thus the iteration bound gives an FPTAS for equal earnings, but
extending it to arbitrary earning requirements remains open.
\end{remark}

\section{Computing a \texorpdfstring{$(3+\eps)$}{(3+epsilon)}-EFX Allocation}\label{sec:three-efx}

In this section, we prove \cref{thm:main} by converting an approximate ER equilibrium into an approximate EFX allocation. Our main result is the following.

\begin{theorem}\label{thm:conversion}
Given a $\delta$-approximate ER equilibrium with unit earning
requirements and earning limits $1/2$, where $\delta\in(0,1/10]$,
we can compute a $(3+10\delta)$-EFX allocation in time polynomial in the input length and $1/\delta$.
\end{theorem}

We use the \textit{chore-swap framework} of \citet{GM26}, which begins with a suitable initial allocation satisfying a certain condition called $\lambda$-\textit{EFX-friendliness}, for parameter $\lambda\ge1$. It iteratively resolves the envy of agents by performing chore swaps:
an agent who violates $\lambda$-EFX exchanges one designated chore from its bundle, called a \textit{swap chore}, for a least-disutility bundle while retaining her other chores. The friendliness conditions
ensure that each swap establishes $\lambda$-EFX for that agent and
preserves this guarantee for agents who are already processed. Thus the
framework computes a $\lambda$-EFX allocation in polynomial time.

Our task is therefore to construct an EFX-friendly starting allocation.
\citet{GMQ25} show that an exact ER equilibrium can be rounded to an
integral allocation, which is subsequently modified to yield a
$4$-EFX-friendly allocation. There are two challenges in extending this approach. First, our FPTAS gives only an approximate ER
equilibrium. Second, the reallocation must improve upon factor four. We address the first challenge by extending the analysis of their rounding algorithm to approximate earnings. Unfortunately, the re-allocation step has a concrete bottleneck at factor $4$, even with an exact ER equilibrium. We give a new algorithm that constructs a $(3+\eps)$-EFX-friendly allocation by choosing the swap chores and their recipients \textit{flexibly} based on the structure of the rounded allocation. 
This flexibility allows us to satisfy stronger EFX-friendly conditions and obtain our improved guarantee.

% This approach uses a fixed "threshold" to designate all chores of price higher the threshold to be swap chores. Our key insight is to devise an algorithm that chooses this threshold -- and hence the swap chores -- flexibly as determined by the rounded allocation. Some high priced chores can then rejoin their owners' retained bundles, allowing us to satisfy stronger EFX-friendly conditions and obtain our improved guarantee of $(3+\eps)$-EFX.

% When $m\le2n$, an exact EFX allocation can be computed in polynomial time~\citep{KMS23,GMQ25}. 
% We therefore assume $m>2n$. Let $(p,b)$
% be a $\delta$-approximate ER equilibrium with unit earning requirements and earning limits $1/2$, where $\delta\in[0,1/10]$.
% For $\delta>0$, our FPTAS computes such an equilibrium since $m/2>n$. We allow $\delta=0$ in the analysis to include exact ER equilibria. Write $\widehat e_i=\sum_jb_{ij}$ for agent $i$'s earning,
% so $1-\delta\le\widehat e_i\le1+\delta$.

\subsection{The Chore-Swap Framework}\label{sec:friendly-framework}

We overview the chore-swap framework of \cite{GM26}. It starts from a $\lambda$-EFX-friendly allocation, which has two types of agents, active and inactive.
The \textit{inactive} agents are already $\lambda$-envy-free.
Each \textit{active} agent has one \textit{swap chore}, while her
other chores form a \textit{residual bundle} of small disutility.
If she violates $\lambda$-EFX, she can exchange her swap chore for
a least-disutility bundle while retaining her residual bundle.
The factor $\lambda-1$ in the definition below leaves room for this
exchange: if the residual costs at most $(\lambda-1)d_i(B)$, adding
bundle $B$ gives total disutility at most $\lambda d_i(B)$.
The definition imposes the comparisons that allow the swaps to be
performed while preserving the fairness guarantees of other agents.

% \begin{samepage}
\begin{definition}[$\lambda$-EFX-friendly allocation, \citealp{GM26}]\label{def:efx-friendly}
Let $\lambda\ge1$ and let $Y$ be an allocation with nonempty bundles.
The allocation $Y$ is $\lambda$-EFX-friendly if there are a partition
$N=N_0\sqcup N_H$ and a \textit{swap chore}
$j_i\in\argmax_{j\in Y_i}d_{ij}$ for each $i\in N_H$ such that, writing
$R_i=Y_i\setminus\{j_i\}$ for agent $i$'s
\textit{residual bundle},
\begin{align*}
 d_i(Y_i)&\le\lambda d_i(Y_k)
 &&(i,k\in N_0),\\
 d_i(Y_i)&\le\lambda d_i(j_h)
 &&(i\in N_0,\ h\in N_H),\\
 d_i(R_i)&\le(\lambda-1)d_i(Y_k)
 &&(i\in N_H,\ k\in N_0),\\
 d_i(R_i)&\le(\lambda-1)d_i(j_h)
 &&(i,h\in N_H).
\end{align*}
\end{definition}
% \end{samepage}

Here $N_H$ and $N_0$ are the sets of active and inactive agents,
respectively. The first two inequalities imply that each inactive
agent is $\lambda$-envy-free. The last two bound each active agent's
residual disutility against every inactive bundle and every swap
chore. An active agent need not perform a swap if she already
satisfies $\lambda$-EFX.

\begin{theorem}[\citealp{GM26}]\label{thm:friendly-conversion}
Given a $\lambda$-EFX-friendly allocation, a $\lambda$-EFX allocation
can be computed in polynomial time using at most $n$ chore swaps.
\end{theorem}

The algorithm first sets aside the swap chores and lets the active
agents pick their least-disutility remaining chore in an arbitrary
fixed order. It then processes these agents in the same order.
An agent who violates $\lambda$-EFX exchanges her swap chore for
a least-disutility bundle, keeping her residual bundle; the other
agent receives the swap chore as a singleton. The picking order
ensures that later swaps preserve the $\lambda$-EFX guarantees of
agents already processed. For completeness, we present this algorithm and sketch its analysis in \cref{app:framework}.

Thus, to prove \cref{thm:conversion}, it
suffices to construct a $(3+10\delta)$-EFX-friendly allocation. Our approach is to round a $\delta$-approximate ER equilibrium, and transform the rounded solution into a $(3+10\delta)$-EFX-friendly allocation. We discuss these algorithms in the following sections.  

\subsection{Rounding an Approximate ER Equilibrium}\label{sec:rounding}

When $m\le2n$, an exact EFX allocation can be computed in polynomial time~\citep{KMS23,GMQ25}. We therefore assume $m>2n$. Let $(p,b)$
be a $\delta$-approximate ER equilibrium with unit earning requirements and earning limits $1/2$, where $\delta\in[0,1/10]$.
For $\delta>0$, \cref{alg:er} computes such an equilibrium since $m/2>n$.

We show how to round an approximate ER equilibrium
into an integral allocation in polynomial time with price bounds that will allow us to
construct an EFX-friendly allocation. 
\citet{GMQ25} give such an algorithm for rounding an \textit{exact}
ER equilibrium $(p,b)$. We first describe their algorithm and its guarantees.
% We present the algorithm using prices and payments $(p,b)$, where
% $b_{ij}=p_jx_{ij}$, in place of allocations and prices $(x,p)$.
Let $L=\{j:p_j\le1/2\}$ and $H=\{j:p_j>1/2\}$ denote the sets of
\textit{low-price} and \textit{high-price} chores, respectively.

\paragraph{The rounding algorithm.}
Define the payment graph of $(p,b)$ as the bipartite graph on $N\sqcup M$
with an edge $(i,j)$ of weight $b_{ij}$ whenever $b_{ij}>0$.
Cancel alternating cycles in the payment graph by subtracting the minimum weight on one
alternating set of edges and adding it on the other. Each cancellation
preserves agent earnings and chore payouts and removes at least one
edge. Thus polynomially many cancellations give a forest while
preserving the ER conditions. Root each tree at an arbitrary agent.
All references to parents and children use this rooted forest.

The algorithm starts with empty integral bundles and proceeds in
four phases, as specified in \cref{alg:rounding}.
\emph{Phase 1} (Line~\ref{line:rounding-leaves}) assigns every original leaf chore to its only adjacent
agent. \emph{Phase 2} (Lines~\ref{line:rounding-low-start}--\ref{line:rounding-low-end}) processes agents in breadth-first order and
allocates all low-price chores, using a cutoff of one for each
agent's current bundle price. An agent whose bundle price already exceeds
one passes her child high-price chores to child agents as well.
\emph{Phase 3} (Line~\ref{line:rounding-prune}) retains exactly one child-agent edge for each remaining
high-price chore, choosing a child who earns the most from that chore.
Every remaining chore then has degree two.
\emph{Phase 4} (Lines~\ref{line:rounding-match-start}--\ref{line:rounding-match-end}) matches these chores to all but one agent in each
remaining tree; the unmatched agent is one with largest current
bundle price.

To obtain the matching in Phase 4, contract each chore to an edge
between its two adjacent agents, root this tree at the agent to be
left unmatched, and assign each edge to its endpoint farther from
the root. All assignments use edges of the original payment graph,
so the resulting allocation is MPB. 
% The rounding procedure and its cutoff at one are unchanged from \citet{GMQ25}.
For the rounded allocation $X$, let $k_i=|X_i\cap H|$ and
$S_i=X_i\setminus H$. When the input is an exact ER equilibrium,
the analysis of \citet{GMQ25} gives, for every agent $i$,
\begin{equation}\label{eq:exact-rounding}
 p(X_i)\ge\frac12,\qquad k_i\le2,\qquad
 p(S_i)\le\frac32-\frac{k_i}{2}.
\end{equation}

\begin{algorithm}[t]
\caption{ER rounding, adapted from \citet{GMQ25}}\label{alg:rounding}
\textbf{Input:} An approximate ER equilibrium $(p,b)$ with unit earnings and earning limits $1/2$\\
\textbf{Output:} An integral MPB allocation $X$
\begin{algorithmic}[1]
\State Cancel cycles in the payment graph; root every resulting tree at an agent
\State $X_i\gets\emptyset$ for every $i$; let $L=\{j:p_j\le1/2\}$ and $H=M\setminus L$
\State Assign each leaf chore to its only adjacent agent and delete it\label{line:rounding-leaves}
\For{each agent $i$, in breadth-first order in the original rooted forest}\label{line:rounding-low-start}
\If{$p(X_i)>1$}
\For{each remaining child chore $j\in H$ of $i$}
\State Assign $j$ to a child maximizing $b_{hj}$; delete $j$
\EndFor
\EndIf
\While{some remaining child $j\in L$ satisfies $p(X_i)+p_j\le1$}
\State Assign $j$ to $i$; delete $j$
\EndWhile
\State Assign every remaining child chore $j\in L$ to any child agent of $j$; delete $j$\label{line:rounding-low-end}
\EndFor
\State For each remaining chore, retain its parent edge and one child edge maximizing $b_{hj}$\label{line:rounding-prune}
\For{each remaining tree $T$}\label{line:rounding-match-start}
\State Choose $h\in N(T)$ maximizing $p(X_h)$
\State Match its chores to $N(T)\setminus\{h\}$ along retained edges, and assign them\label{line:rounding-match-end}
\EndFor
\State \Return $X$
\end{algorithmic}
\end{algorithm}

\paragraph{Approximate ER equilibria.}
Our FPTAS, however, only gives a $\delta$-approximate ER equilibrium,
in which each agent's earning lies in $[1-\delta,1+\delta]$.
Even an arbitrarily small earning error can allow an agent to receive
\textit{three} high-price chores, as the following example shows.

\begin{example}[A third high-price chore]\label{ex:third-high}
Consider three agents with identical disutilities equal to the chore
prices. There are four high-price chores, each of price $P>2$.
In the ER flow, two are supported only by agent 1, one only by
agent 2, and the fourth pays $\delta$ to agent 1 and $1/2-\delta$
to agent 2, where $0<\delta<1/2$. Agent 3 alone performs three
low-price chores of price $1/3$ each. The earnings are
$(1+\delta,1-\delta,1)$, and all approximate ER conditions hold.

Root the shared component at agent 2. Phase 1 of
\cref{alg:rounding} assigns each leaf chore to its only adjacent agent. Agent 2 then
holds a chore of price greater than one, so the procedure passes
her shared high-price chore to its child agent, agent 1.
Thus agent 1 receives three high-price chores even for an
arbitrarily small earning error $\delta>0$.
\end{example}

We therefore need a new analysis of the rounding algorithm for
approximate earnings. The useful guarantee is a joint bound on the
number of high-price chores and the total price of low-price chores.
In particular, an agent who receives three high-price chores must
have $p(S_i)\le\delta$. We prove the following.

\begin{lemma}[Rounding with approximate earnings]\label{lem:approx-rounding}
Let $\delta\in[0,1/2)$. Given a $\delta$-approximate ER equilibrium
with unit earnings and earning limits $1/2$,
\cref{alg:rounding} returns an integral MPB allocation $X$ in
polynomial time. For every agent $i$, with $k_i = |X_i\cap H|$ and $S_i = X_i\setminus H$, we have
\begin{equation}\label{eq:approx-rounding}
 p(X_i)\ge\frac12-\delta,\qquad k_i\le3,\qquad
 p(S_i)\le\frac32+\delta-\frac{k_i}{2}.
\end{equation}
\end{lemma}

\begin{proof}[Proof of \cref{lem:approx-rounding}]
MPB optimality and polynomial running time follow from the
algorithm. We prove the price bounds using the acyclic payment graph
before Phase~1. Write $\widehat e_i=\sum_jb_{ij}$ for agent $i$'s
original earning, so $1-\delta\le\widehat e_i\le1+\delta$.

\paragraph{Upper bounds.}
Let $F_i$ be the original leaf chores assigned to agent $i$ in
Phase 1. Since a low-price leaf pays its full price and a high-price
leaf pays $1/2$, we have
\begin{equation}\label{eq:leaf-budget}
 p(F_i\cap L)+\frac12|F_i\cap H|
 \le\widehat e_i\le1+\delta.
\end{equation}
In particular, $F_i$ contains at most two high-price chores.
When $i$ is visited in Phase 2, her current bundle consists of $F_i$
and possibly her original parent chore.

If this bundle has price at most one, the low-price assignments keep
it at most one. Agent $i$ may subsequently receive at most one
high-price chore in Phase 4. A bundle of price at most one contains
at most one high-price chore, since such chores have price strictly
greater than $1/2$. Therefore $k_i\le2$, and
$p(S_i)\le3/2-k_i/2$.

Otherwise the bundle price at $i$'s visit exceeds one. All her child
chores are passed down or removed. Thus her final bundle consists
of $F_i$ and at most her original parent chore, whether that parent
was assigned earlier or in Phase 4. If the parent is low-price,
\eqref{eq:leaf-budget} gives
$p(S_i)\le1+\delta-k_i/2+1/2$. If it is high-price and assigned to
$i$, it contributes one to $k_i$ and nothing to $S_i$, giving
$p(S_i)\le1+\delta-(k_i-1)/2$. If the parent is not assigned to
$i$, the bound is stronger. In every case $k_i\le3$ and the
upper bound in \eqref{eq:approx-rounding} follows.

When $\delta=0$, two high-price original leaves consume the entire
unit earning of $i$. Such an agent cannot have a positive parent
edge, so cannot obtain a third high-price chore. This gives $k_i\le2$
in the exact case.

\paragraph{Lower bound.}
Consider a Phase-3 tree $T$, and let $h$ be an agent of maximum
current bundle price. Every other agent receives a high-price chore
in Phase 4, so obtains price greater than $1/2$. It suffices to
lower-bound $p(X_h)$, which equals $h$'s price before Phase 4.

Suppose some agent $i\in N(T)$ had a child chore assigned to another
agent in Phase 2. If that chore was high-price, her bundle already
had price greater than one. If it was low-price, it did not fit under
the cutoff one, so her bundle price was greater than $1-1/2=1/2$.
Bundle prices only increase during rounding. By the choice of $h$,
we therefore have $p(X_h)\ge1/2$ in this case.

Otherwise no agent in $N(T)$ lost a child chore in Phase 2. All
original earnings of these agents are represented either by chores
already assigned to them or by the $|N(T)|-1$ high-price chores
remaining inside $T$, except possibly for the original parent chore
of the root of $T$. This exceptional chore accounts for at most
$1/2$ of the root's original earning. Every internal high-price
chore also pays out $1/2$. Since every agent originally earned at
least $1-\delta$, the sum of their current bundle prices is at least
\[
 |N(T)|(1-\delta)-\frac12
             -\frac12\bigl(|N(T)|-1\bigr)
 =|N(T)|\left(\frac12-\delta\right).
\]
Therefore some agent in $T$, and hence also the maximum-price agent $h$, has bundle price at least $1/2-\delta$. This proves the
lower bound.
\end{proof}

We will use these rounding bounds in \cref{sec:threshold} to
construct a $(3+10\delta)$-EFX-friendly allocation for the chore-swap
framework. As a warm-up, we recall in the next section the known construction of
a $4$-EFX-friendly allocation from an exact ER equilibrium.

\subsection{Obtaining a 4-EFX-Friendly Allocation}\label{sec:four-efx}

We present the construction of a $4$-EFX-friendly allocation from an ER equilibrium due to \citet{GMQ25} in the language of EFX-friendliness from \citet{GM26}.
Let $(X,p)$ be the allocation and prices obtained by rounding an
exact ER equilibrium with unit earnings and
earning limits $1/2$. Thus $X$ satisfies \eqref{eq:exact-rounding}.
Normalize disutilities by the MPB ratios of $X$, as in
\eqref{eq:mpb-normalization}. Each agent's own chores then have
disutility equal to their price, and every chore has disutility
at least its price to every agent.

The construction sets aside all high-price chores $H$ and reallocates
them, leaving each agent's low-price chores with her. Agents who
receive several high-price chores are shown to be sufficiently
envy-free and are inactive. Each active agent receives exactly one
high-price chore as her swap chore and retains $S_i=X_i\setminus H$
as her residual bundle. Thus a \textit{fixed} threshold of $1/2$
determines which chores are reallocated and provides the lower bound
on swap-chore prices used in the friendliness comparisons.

\begin{lemma}[\citealp{GM26}]\label{lem:fixed-friendly}
Given an allocation $(X,p)$ obtained by rounding an exact ER
equilibrium with unit earning requirements and earning limits $1/2$,
a $4$-EFX-friendly allocation can be constructed in polynomial time.
\end{lemma}
\begin{proof}
We use the bounds in \eqref{eq:exact-rounding}. In particular,
$|H|\le2n$, and every agent's low-price chores have total price at
most $3/2$.

If $|H|\le n$, assign one high-price chore to each of $|H|$ agents,
including every original owner of a high-price chore, while retaining
all low-price chores. Denote the resulting allocation by $Y$.
The high-price chore recipients form $N_H$; all other agents are
inactive and keep $Y_i=X_i=S_i$.
Every inactive agent has disutility at most $3/2$ for her bundle,
while every inactive bundle and every swap chore has price at least
$1/2$. Thus the first two conditions of \cref{def:efx-friendly}
hold even with factor three.

If $n<|H|\le2n$, apply the few-chores EFX algorithm of
\citet{GMQ25} to $H$, ordering the original owners of high-price
chores first. There are at least $|H|-n$ such agents, since each
originally owns at most two high-price chores and $|H|\le2n$.
The algorithm returns an EFX allocation $Z$ with nonempty bundles,
where only the first $|H|-n$ agents can receive several chores.
Set $Y_i=S_i\cup Z_i$, let $N_H=\{i:|Z_i|=1\}$, and let
$N_0=N\setminus N_H$. For every inactive agent $i$, we have
$k_i\ge1$, and hence $p(S_i)\le1$. Let $g$ be a least-disutility
chore in $Z_i$. Since $|Z_i|\ge2$ and $Z$ is EFX,
\[
 d_i(Z_i)\le2d_i(Z_i\setminus\{g\})\le2d_i(Z_h)
\qquad(h\in N).
\]
Consequently,
\[
 d_i(Y_i)=d_i(S_i)+d_i(Z_i)
\le1+2d_i(Z_h)\le4d_i(Z_h),
\]
since $d_i(Z_h)\ge p(Z_h)>1/2$. When $h$ is active, $Z_h$ is
its singleton swap chore; when $h$ is inactive, $Z_h\subseteq Y_h$.
This proves the first two friendliness conditions.

In both cases, every active agent retains a residual bundle $S_i$
of disutility at most $3/2$. Every inactive bundle and every swap
chore has price at least $1/2$, so the last two friendliness
conditions hold with factor $\lambda-1=3$. Each swap chore is also
a highest-disutility chore for its recipient: its disutility is
at least its price, which exceeds $1/2$, whereas every residual
chore has disutility equal to its price and at most $1/2$.
All bundles are nonempty, and the reallocation takes polynomial time.
Thus the resulting allocation is $4$-EFX-friendly.
\end{proof}

Together with \cref{thm:friendly-conversion}, this recovers the
$4$-EFX guarantee of \citet{GMQ25} in polynomial time given an
exact ER equilibrium as input.

To obtain our $(3+\eps)$-EFX guarantee, we must address two obstacles.
First, rounding an approximate ER equilibrium can give an agent
three high-price chores, as shown in \cref{ex:third-high}. The
argument above uses the two-chore bound when choosing recipients
for the reallocation. More importantly, even with an exact ER equilibrium,
this fixed-threshold construction has a bottleneck at factor four.
A residual bundle of an active agent has price at most $3/2$, while a swap chore can
have price arbitrarily close to $1/2$. Comparing these quantities
in the friendliness condition gives $\lambda-1=3$, and hence
$\lambda=4$. The following example shows that this loss occurs
in an actual ER rounding.

\begin{example}[A bottleneck in the fixed-threshold construction]\label{ex:fixed-threshold}
Consider four agents with identical disutilities. Fix $0<\beta<1/6$
and let $a=(1-\beta)/2$ and $p=1/2+\beta$. Suppose the rounded
allocation is $X_1=\{a,a,1/2\}, X_2=\{\beta,p\},
 X_3=X_4=\{p,p\},$ where each entry denotes the price of a distinct chore, equal to
every agent's disutility for that chore.

The fixed threshold is $1/2$: the construction in
\cref{lem:fixed-friendly} sets aside all five chores of price
$p>1/2$ for reallocation, while each agent retains her low-price
chores. In particular, agent 1 retains $S_1=X_1$, of disutility
$3/2-\beta$. The reallocation gives two high-price chores to one
agent and one to each other agent. Agents 2, 3, and 4 are ordered
before agent 1 because they originally own high-price chores.
Only the first agent in this order can receive two chores, so
agent 1 receives one high-price chore and becomes active.
Every swap chore has disutility $p$, and hence EFX-friendliness
requires $3/2-\beta\le(\lambda-1)p$, or
\[
 \lambda\ge1+\frac{3/2-\beta}{p}
 =\frac{4}{1+2\beta}
\]
% \longrightarrow4 \qquad(\beta\longrightarrow0).
This approaches factor $4$ as $\beta$ approaches $0$.
Thus this construction cannot guarantee $\lambda$-EFX-friendliness
for any fixed $\lambda<4$.

To avoid this loss, we allow the threshold to increase from $1/2$
to $t=p$. Starting again from $X$, set aside the five high-price
chores as candidate swap chores. At the new threshold 
$p$, let one high-price chore rejoin the residual bundle of each
of agents 3 and 4. Now three candidate swap chores remain, one originally owned
by each of agents 2, 3, and 4. Keep these as their swap chores
and leave agent 1 inactive. Thus $N_0=\{1\}$, $N_H=\{2,3,4\}$,
and the active residual bundles are $R_2=\{\beta\}$ and
$R_3=R_4=\{p\}$. We have
\[
 t\le d_1(X_1)=3/2-\beta\le3t,
 \qquad d_i(R_i)\le2t\quad(i\in N_H).
\]
Every swap chore has disutility $t$ and is a highest-disutility
chore in its owner's bundle. Since the agents have identical
disutilities, these bounds verify all four conditions of
\cref{def:efx-friendly} for $\lambda=3$. Hence $X$ itself is
$3$-EFX-friendly. Raising the threshold lets agents 3 and 4 retain
one high-price chore each in their residual bundles, so agent 1
can remain inactive despite her large residual bundle.
\end{example}

\subsection{Constructing an EFX-Friendly Allocation}\label{sec:threshold}

\Cref{ex:fixed-threshold} illustrates how a \textit{flexible} threshold can
overcome the factor-$4$ bottleneck: allowing some high-price chores
to rejoin their owners' residual bundles lets agents with large
residual bundles remain inactive. We generalize this idea to the allocation
$(X,p)$ obtained by rounding a $\delta$-approximate ER equilibrium
with unit earning requirements and earning limits $1/2$. Our
construction uses the joint bounds in \cref{lem:approx-rounding},
which allow an agent to hold three high-price chores.

We begin with a threshold $t=1/2-\delta$ and consider all high-price chores to be \textit{candidate swap chores}.
Our algorithm iteratively increases the price threshold $t$
and reduces the set of candidate swap chores, letting some candidate chores
rejoin their original owners' residual bundles. We choose the threshold and the
remaining candidate chores so that each active agent can receive one swap
chore, while the other agents keep their residual bundles. The sets
$L$ and $H$ remain defined by the fixed cutoff $1/2$ throughout.

\paragraph{The required price bounds.}
At any stage of the construction, let $E\subseteq H$ denote
the current set of candidate swap chores and $R_i=X_i\setminus E$
the residual bundle of agent $i$.
Normalize disutilities using the MPB ratios of $X$. Since residual
chores remain with their original owners, $d_i(R_i)=p(R_i)$, while
any other bundle has disutility at least its price. To obtain a
$\lambda$-EFX-friendly allocation, for a factor $\lambda\ge3$, we
seek a partition $N=N_0\sqcup N_H$, with $|N_H|=|E|$, satisfying
\begin{equation}\label{eq:friendly-price-bounds}
\begin{aligned}
 t\le p(R_i)&\le\lambda t &&(i\in N_0),\\
 p(R_i)&\le(\lambda-1)t &&(i\in N_H),\\
 p_j&\ge t &&(j\in E).
\end{aligned}
\end{equation}
Each active agent then receives one distinct chore from $E$.
These bounds give the four disutility comparisons in
\cref{def:efx-friendly}. We will maintain the upper bound
$p(R_i)\le\lambda t$ for every agent throughout the construction.

We must also ensure that each swap chore is a highest-disutility
chore in its recipient's bundle. For this reason, we return a
candidate chore to its owner only when $t$ reaches its price. Every
returned chore then has price at most that of any remaining
candidate chore. The original low-price chores also have smaller prices
than all candidate chores. Since residual chores remain on MPB assignments,
these price comparisons ensure the required disutility ordering.

\paragraph{Choosing recipients for the swap chores.}
For a given threshold $t$ and candidate set $E$, write
\[
 u_i=p(R_i),\qquad
 A=\{i:u_i\le(\lambda-1)t\},\qquad D=\{i:u_i<t\}.
\]
The agents in $A$ are \textit{eligible}: their residual bundles
are small enough for them to receive a swap chore. The agents in
$D$ are \textit{deficient}: they must receive a swap chore, since
otherwise their bundles would have price below $t$. Hence we need
$D\subseteq N_H\subseteq A$ and $|N_H|=|E|$. Such a choice exists
precisely when
\[
 |D|\le|E|\le|A|.
\]
The two inequalities express the competing requirements: there must
be enough swap chores for deficient agents, and enough eligible
agents for the swap chores.

\paragraph{The threshold procedure.}
We start with $E=H$ and $R_i=S_i$. To ensure that there are initially
enough candidate chores for all deficient agents, set $t=1/2-\delta$,
the lower bound on rounded bundle prices. Indeed, if $p(S_i)<1/2-\delta$,
then $X_i$ must contain a high-price chore because $p(X_i)\ge1/2-\delta$.
Thus each initially deficient agent owns a distinct candidate chore.
This choice of $t$ uses the lower bound on whole bundles; the
candidate set is still $H=\{j:p_j>1/2\}$.

Initially, the residual prices satisfy $p(S_i)\le3/2+\delta$.
To ensure $u_i\le\lambda t$ at the start, set
\begin{equation}\label{eq:parameters}
 \lambda=\frac{3/2+\delta}{1/2-\delta}=\frac{3+2\delta}{1-2\delta}.
\end{equation}
Since $\delta\ge0$, we have $\lambda\ge3$.
For an exact ER input, $\lambda=3$, giving the bounds $3t$ for
inactive bundles and $2t$ for active residual bundles seen in
\cref{ex:fixed-threshold}. We will use the joint rounding bound to
show that $u_i\le\lambda t$ continues to hold as chores return
to their owners' residual bundles.

While $|E|>|A|$, there are too many candidate chores for the eligible
agents. We either return a candidate chore of price $t$ to its original
owner or, if none has this price, increase $t$. A threshold increase
stops at the next candidate price or the next value $u_i/(\lambda-1)$ at which
an ineligible agent becomes eligible, whichever is smaller.
Reaching a candidate price allows a chore to be returned.
Stopping at an eligibility event ensures that an agent outside $A$
cannot become deficient during a single threshold increase: she
becomes eligible at $t=u_i/(\lambda-1)$, before becoming deficient at $t>u_i$.

We stop as soon as $|E|\le|A|$. Chores of the same price are
returned one at a time, checking this condition after every event.
The key observation, proved below, is that these rules preserve
$|D|\le|E|$. Thus, when there are enough eligible recipients, there
are still enough swap chores for all deficient agents.
\Cref{alg:threshold} gives the algorithm.

\begin{algorithm}[t]
\caption{Constructing an EFX-friendly allocation}\label{alg:threshold}
\textbf{Input:} A rounded MPB allocation $(X,p)$ satisfying \eqref{eq:approx-rounding}\\
\textbf{Output:} A $\lambda$-EFX-friendly allocation $Y$, its partition, and swap chores
\begin{algorithmic}[1]
\State Set $\lambda$ as in \eqref{eq:parameters}
\State $t\gets1/2-\delta$; $E\gets H$; $R_i\gets X_i\setminus H$ for every $i$
\State $u_i\gets p(R_i)$; $A\gets\{i:u_i\le(\lambda-1)t\}$; $D\gets\{i:u_i<t\}$
\While{$|E|>|A|$}
\If{some $j\in E$ has $p_j=t$}
\State Choose one such $j$ and let $i$ be its owner in $X$
\State $E\gets E\setminus\{j\}$; $R_i\gets R_i\cup\{j\}$; $u_i\gets u_i+p_j$
\Else
\State $t\gets\min\bigl(\{p_j:j\in E\}\cup\{u_i/(\lambda-1):i\notin A\}\bigr)$
\EndIf
\State $A\gets\{i:u_i\le(\lambda-1)t\}$; $D\gets\{i:u_i<t\}$
\EndWhile
\State Choose $N_H$ with $D\subseteq N_H\subseteq A$ and $|N_H|=|E|$
\State $N_0\gets N\setminus N_H$; assign one distinct $j_i\in E$ to each $i\in N_H$
\State $Y_i\gets R_i\cup\{j_i\}$ for $i\in N_H$; $Y_i\gets R_i$ for $i\in N_0$
\State \Return $Y$, $N_0$, $N_H$, and the swap chores $j_i$
\end{algorithmic}
\end{algorithm}

The candidate set $E$ is nonempty whenever the while-condition holds.

\begin{samepage}
\begin{lemma}[Properties of the threshold procedure]\label{lem:threshold}
\Cref{alg:threshold} terminates after $O(n+|H|)$ events. At termination,
\begin{equation}\label{eq:threshold-certificate}
 |D|\le|E|\le|A|,\qquad
 u_i\le\lambda t\quad(i\in N),\qquad
 p_j\ge t\quad(j\in E).
\end{equation}
Each $R_i$ consists of $S_i$ and returned high-price chores
originally assigned to $i$ in $X$, each of price at most $t$.
\end{lemma}
\end{samepage}
\begin{proof}
By construction, residual chores remain with their original
owners. Candidate chores have price at least $t$, and returned high-price
chores have price at most $t$. Chores of price exactly $t$ may lie
in either set when the procedure stops.

\paragraph{Residual prices.}
Before $t$ reaches $1/2$, no high-price chore can be returned, so
\[
 u_i=p(S_i)\le\frac32+\delta=\lambda\left(\frac12-\delta\right)\le\lambda t.
\]
For $t\ge1/2$, returned chores increase the residual price. However,
the upper bound on $p(S_i)$ decreases by $1/2$ for each high-price
chore originally assigned to $i$. By \cref{lem:approx-rounding}
and $k_i\le3$,
\[
\begin{aligned}
 u_i
 &\le\frac32+\delta-\frac{k_i}{2}+k_it\\
 &\le3t+\delta\le\lambda t.
\end{aligned}
\]
The second inequality uses $t\ge1/2$ and $k_i\le3$; the last uses
$\lambda-3=8\delta/(1-2\delta)\ge2\delta$ and $t\ge1/2$.
Thus the joint rounding bound accommodates even a third high-price
chore returning to its owner's residual bundle.

\paragraph{Enough swap chores for deficient agents.}
Initially, every agent with $p(S_i)<1/2-\delta$ must own a high-price chore,
since $p(X_i)\ge1/2-\delta$. Choosing one such owned chore per deficient
agent gives $|D|\le|H|=|E|$.
Before every nonterminal event,
\begin{equation}\label{eq:strict-count}
 |D|\le|A|<|E|,
\end{equation}
because $D\subseteq A$. Returning one chore decreases $|E|$ by one
and can only shrink $D$, since $t$ stays fixed and the owner's
residual price increases. The strict inequality in
\eqref{eq:strict-count} therefore gives $|D|\le|E|$ after the
return, even if the owner leaves $A$.

Suppose instead that the threshold increases to $t'$. By the choice
of the next event, every agent outside the old $A$ has
$u_i\ge(\lambda-1)t'\ge t'$, and hence cannot belong to the new $D$.
Therefore
\[
 D_{\mathrm{new}}\subseteq A_{\mathrm{old}},\qquad
 |D_{\mathrm{new}}|\le|A_{\mathrm{old}}|<|E|.
\]
This proves $|D|\le|E|$ after each event, including the first one
at which the stopping condition $|E|\le|A|$ holds.

\paragraph{Number of events.}
At most $|H|$ chores are returned. Between returns, increasing $t$
can only add agents to $A$. Each return removes at most one agent
from $A$, so at most $n+|H|$ entries into $A$ can occur. Every
threshold increase either adds an eligible agent or reaches a
candidate price, in which case a chore is returned next unless
the procedure stops. Thus there are $O(n+|H|)$ events. If $E$
becomes empty, the stopping condition holds, so the procedure
terminates.
\end{proof}

We can now choose the active agents and assign the remaining swap
chores. The price bounds imply the required disutility comparisons.

\begin{lemma}\label{lem:threshold-friendly}
Given the rounded allocation in \cref{lem:approx-rounding},
\cref{alg:threshold} constructs a $\lambda$-EFX-friendly allocation
in polynomial time, for $\lambda$ in \eqref{eq:parameters}.
For an exact ER input, the resulting allocation is $3$-EFX-friendly.
\end{lemma}
\begin{proof}
By \cref{lem:threshold}, the algorithm can choose $N_H$ with
$D\subseteq N_H\subseteq A$ and $|N_H|=|E|$. In the resulting
allocation $Y$, every active agent receives one swap chore and
every inactive agent keeps her residual bundle. The definitions
of $A,D$, together with $u_i\le\lambda t$, give
\[
 t\le p(R_i)\le\lambda t\quad(i\in N_0),\qquad
 p(R_i)\le(\lambda-1)t\quad(i\in N_H).
\]
Thus \eqref{eq:friendly-price-bounds} holds. Since retained chores
were originally assigned on MPB edges, $d_i(R_i)=p(R_i)$. Every
inactive bundle and every swap chore has price at least $t$, so
\begin{align}
 d_i(Y_i)&\le\lambda d_i(B)
 &&(i\in N_0),\label{eq:protected}\\
 d_i(R_i)&\le(\lambda-1)d_i(B)
 &&(i\in N_H),\label{eq:active}
\end{align}
for every $B\in\{Y_k:k\in N_0\}\cup\{\{j_h\}:h\in N_H\}$.
These are the four inequalities in \cref{def:efx-friendly}.

It remains to check that each swap chore $j_i$ is a
highest-disutility chore in its recipient's bundle. By
\cref{lem:threshold}, every $g\in R_i$ has price at most
$\max\{1/2,t\}$, whereas $p_{j_i}>1/2$ and $p_{j_i}\ge t$.
Hence
\[
 d_i(g)=p_g\le p_{j_i}\le d_i(j_i).
\]
Inactive bundles have price at least $t>0$, and active agents each
receive a swap chore, so all bundles are nonempty. This proves
friendliness. The construction takes polynomial time by
\cref{lem:threshold}; when $\delta=0$, the factor is $\lambda=3$.
\end{proof}

\subsection{The Approximation Guarantee and Running Time}

We are now in a position to complete the proof of \cref{thm:conversion}, and that of the main result of our paper, \cref{thm:main}.

\begin{proof}[\bf Completing the proof of \cref{thm:conversion}]
Apply \cref{alg:rounding} and then \cref{alg:threshold}.
By \cref{lem:approx-rounding,lem:threshold-friendly}, this gives a
$\lambda$-EFX-friendly allocation for
$\lambda=(3+2\delta)/(1-2\delta)$. By
\cref{thm:friendly-conversion}, at most $n$ chore swaps yield a
$\lambda$-EFX allocation.

Since $\delta\le1/10$, its approximation factor satisfies
\[
 \lambda=\frac{3+2\delta}{1-2\delta}
 =3+\frac{8\delta}{1-2\delta}\le3+10\delta.
\]
Rounding takes polynomial time, and threshold selection uses
$O(n+|H|)$ events. The chore-swap procedure also takes
polynomial time. Hence the full algorithm computes a
$(3+10\delta)$-EFX allocation from the given $\delta$-approximate
ER equilibrium in polynomial time.
\end{proof}

\begin{proof}[\bf Proof of \cref{thm:main}]
For $m\le2n$, use the known polynomial-time exact EFX algorithm \cite{KMS23,GMQ25}.
Otherwise, set $\delta=\eps/10$. 
% Since $\eps\in(0,1)$, we have $0<\delta<1/10$. 
Compute a $\delta$-approximate ER equilibrium with caps $1/2$
using \cref{thm:er-fptas}, and apply \cref{thm:conversion}.
The resulting allocation is $(3+10\delta)$-EFX, and hence
$(3+\eps)$-EFX.

Since $1/\delta=10/\eps$, the number of flow
computations is $O(1+n\log(mR)/\eps^2)$. The remaining steps take
polynomial time by \cref{thm:conversion}. Thus the full running time
is polynomial in the input length and $1/\eps$.
\end{proof}

\section{Conclusions}\label{sec:conclusion}
In this paper, we gave a polynomial-time algorithm for computing a $(3+\eps)$-EFX
allocation of chores with additive disutilities. Our approach
combines an FPTAS for equal-earnings ER equilibria with a novel threshold-based
construction of EFX-friendly allocations. Two natural questions remain: can the polynomial-time EFX
factor be improved to $2+\eps$, matching the known existence guarantee
up to an arbitrarily small loss, and can an ER FPTAS be obtained for
arbitrary, non-identical earning requirements?

\paragraph{Statement of AI Use.}
The results of this paper were obtained in late September
2026; since then the author focused on polishing the exposition. The author used ChatGPT 5.6, 6, and 6.1 in developing the results and preparing this manuscript. The FPTAS for ER equilibria and was developed through interactions with ChatGPT. The author supplied the conceptual connections to Shmyrev's formulation and spending-restricted equilibria, together with the associated intuition and market interpretation to formulate the new program for earning restricted equilibria. The convergence analysis of mirror-ascent was entirely due to ChatGPT. The idea of changing the price threshold for converting an ER rounded allocation to an EFX-friendly one originated during the author's earlier work on \citet{GMQ25} and \citet{GM26}. By choosing a different -- but fixed -- threshold, the author obtained a $(2+\sqrt{3})$-EFX bound. The improved construction using flexible thresholds was developed from this idea through a sequence of conversations with ChatGPT. The author takes full responsibility for the paper.

\bibliographystyle{plainnat}
\bibliography{references}

@article{foleyEF,
  author={Duncan K. Foley},
  title={Resource Allocation and the Public Sector},
  journal={Yale Economic Essays}, volume={7}, number={1}, pages={45--98}, year={1967}
}

@book{mas1995microeconomic,
  title={Microeconomic Theory},
  author={Mas-Colell, A. and Whinston, M.D. and Green, J.R.},
  year={1995},
  publisher={Oxford University Press}
}

@article{Shmyrev09,
  author={V. I. Shmyrev},
  title={An Algorithm for Finding Equilibrium in the Linear Exchange Model with Fixed Budgets},
  journal={Journal of Applied and Industrial Mathematics},
  volume={3}, number={4}, pages={505--518}, year={2009},
  doi={10.1134/S1990478909040097}
}

@article{BeckTeboulle03,
  author={Amir Beck and Marc Teboulle},
  title={Mirror Descent and Nonlinear Projected Subgradient Methods for Convex Optimization},
  journal={Operations Research Letters},
  volume={31}, number={3}, pages={167--175}, year={2003},
  doi={10.1016/S0167-6377(02)00231-6},
  url={https://www.tau.ac.il/~becka/3.pdf}
}

@article{Eisenberg1959EGConvexProgram,
  author={Edmund Eisenberg and David Gale},
  title={Consensus of Subjective Probabilities: The Pari-Mutuel Method},
  journal={The Annals of Mathematical Statistics},
  volume={30}, number={1}, pages={165--168}, year={1959},
  doi={10.1214/aoms/1177706369},
  url={https://www.jstor.org/stable/2237130}
}

@article{bogomolnaia2017manna,
  author={Anna Bogomolnaia and Herv{\'e} Moulin and Fedor Sandomirskiy and Elena Yanovskaya},
  title={Competitive Division of a Mixed Manna},
  journal={Econometrica},
  volume={85}, number={6}, pages={1847--1871}, year={2017},
  doi={10.3982/ECTA14564},
  url={https://arxiv.org/abs/1702.00616}
}

@article{zhou22efx,
  author={Shengwei Zhou and Xiaowei Wu},
  title={Approximately {EFX} Allocations for Indivisible Chores},
  journal={Artificial Intelligence}, volume={326}, pages={104037}, year={2024},
  doi={10.1016/j.artint.2023.104037}
}

@inproceedings{GMQ25,
  author={Jugal Garg and Aniket Murhekar and John Qin},
  title={Constant-Factor {EFX} Exists for Chores},
  booktitle={Proceedings of the 57th Annual ACM Symposium on Theory of Computing},
  pages={1580--1589}, year={2025}, doi={10.1145/3717823.3718305},
  url={https://arxiv.org/abs/2407.03318}
}

@article{GM26,
  author={Jugal Garg and Aniket Murhekar},
  title={Existence of 2-{EFX} Allocations of Chores},
  journal={Proceedings of the AAAI Conference on Artificial Intelligence},
  volume={40}, number={20}, pages={16922--16929}, year={2026},
  doi={10.1609/aaai.v40i20.38738}, url={https://arxiv.org/abs/2507.19461}
}

@misc{QFLLC26,
  author={Genjie Qin and Qizhi Fang and Minming Li and Wenjing Liu and Xin Chen},
  title={Polynomial-Time Approximate {EFX} for Chores via Flow and Matching},
  year={2026}, eprint={2611.07468}, archivePrefix={arXiv},
  primaryClass={cs.GT}, url={https://arxiv.org/abs/2611.07468}
}

@inproceedings{KMS23,
  author={Yusuke Kobayashi and Ryoga Mahara and Souta Sakamoto},
  title={{EFX} Allocations for Indivisible Chores: Matching-Based Approach},
  booktitle={Algorithmic Game Theory}, pages={257--270}, year={2023}
}

@inproceedings{BCM22,
  author={Shant Boodaghians and Bhaskar Ray Chaudhury and Ruta Mehta},
  title={Polynomial Time Algorithms to Find an Approximate Competitive Equilibrium for Chores},
  booktitle={Proceedings of the ACM-SIAM Symposium on Discrete Algorithms},
  pages={2285--2302}, year={2022}, url={https://arxiv.org/abs/2107.06649}
}

@inproceedings{CKMN24,
  author={Bhaskar Ray Chaudhury and Christian Kroer and Ruta Mehta and Tianlong Nan},
  title={Competitive Equilibrium for Chores: From Dual {Eisenberg--Gale} to a Fast, Greedy, {LP}-Based Algorithm},
  booktitle={Proceedings of the ACM Conference on Economics and Computation},
  year={2024}, url={https://arxiv.org/abs/2402.10439}
}

@misc{CJS24,
  author={He Chen and Chonghe Jiang and Anthony Man-Cho So},
  title={Computing Competitive Equilibrium for Chores: Linear Convergence and Lightweight Iteration},
  year={2024}, eprint={2410.04036}, archivePrefix={arXiv},
  url={https://arxiv.org/abs/2410.04036}
}

@inproceedings{ColeEtAl17,
  author={Richard Cole and Nikhil R. Devanur and Vasilis Gkatzelis and Kamal Jain and Tung Mai and Vijay V. Vazirani and Sadra Yazdanbod},
  title={Convex Program Duality, {Fisher} Markets, and {Nash} Social Welfare},
  booktitle={Proceedings of the ACM Conference on Economics and Computation},
  pages={459--460}, year={2017}, url={https://arxiv.org/abs/1609.06654}
}

@misc{HeTao2026,
  author={Wentao He and Biaoshuai Tao},
  title={{EFX} for Additive Chores: Nonexistence, {Pareto} Incompatibility, and Bi-Valued Existence},
  year={2026}, eprint={2606.08872}, archivePrefix={arXiv},
  url={https://arxiv.org/abs/2606.08872}
}

@inproceedings{mahara,
  author={Ryoga Mahara},
  title={Existence of Fair and Efficient Allocation of Indivisible Chores},
  booktitle={Proceedings of the ACM-SIAM Symposium on Discrete Algorithms},
  pages={6742--6766}, year={2026},
  doi={10.1137/1.9781611978971.242},
  url={https://arxiv.org/abs/2507.09544}
}

@misc{AADM24,
  author={Mahyar Afshinmehr and Matin Ansaripour and Alireza Danaei and Kurt Mehlhorn},
  title={Approximate {EFX} and Exact {tEFX} Allocations for Indivisible Chores: Improved Algorithms},
  year={2024}, eprint={2410.18655}, archivePrefix={arXiv},
  url={https://arxiv.org/abs/2410.18655}
}

@inproceedings{LWZ25,
  author={Zehan Lin and Xiaowei Wu and Shengwei Zhou},
  title={Approximately {EFX} and {fPO} Allocations for Bivalued Chores},
  booktitle={Proceedings of the International Joint Conference on Artificial Intelligence},
  pages={3952--3960}, year={2025}, doi={10.24963/ijcai.2025/440},
  url={https://www.ijcai.org/proceedings/2025/440}
}

@article{GMQ22,
  author={Jugal Garg and Aniket Murhekar and John Qin},
  title={Fair and Efficient Allocations of Chores under Bivalued Preferences},
  journal={Proceedings of the AAAI Conference on Artificial Intelligence},
  volume={36}, number={5}, pages={5043--5050}, year={2022},
  doi={10.1609/aaai.v36i5.20436}, url={https://arxiv.org/abs/2110.09601}
}

@inproceedings{EPS22,
  author={Soroush Ebadian and Dominik Peters and Nisarg Shah},
  title={How to Fairly Allocate Easy and Difficult Chores},
  booktitle={Proceedings of the International Conference on Autonomous Agents and Multiagent Systems},
  year={2022}, url={https://arxiv.org/abs/2110.11285}
}

@inproceedings{CGMM22,
  author={Bhaskar Ray Chaudhury and Jugal Garg and Peter McGlaughlin and Ruta Mehta},
  title={Competitive Equilibrium with Chores: Combinatorial Algorithm and Hardness},
  booktitle={Proceedings of the ACM Conference on Economics and Computation},
  pages={1106--1107}, year={2022}, doi={10.1145/3490486.3538255},
  url={https://arxiv.org/abs/2205.11363}
}

@inproceedings{CG15,
  author={Richard Cole and Vasilis Gkatzelis},
  title={Approximating the {Nash} Social Welfare with Indivisible Items},
  booktitle={Proceedings of the ACM Symposium on Theory of Computing},
  pages={371--380}, year={2015}, doi={10.1145/2746539.2746589}
}

@inproceedings{Barman18FFEA,
  author = {Barman, Siddharth and Krishnamurthy, Sanath Kumar and Vaish, Rohit},
  title = {Finding Fair and Efficient Allocations},
  booktitle = {Proceedings of the 19th ACM Conference on Economics and Computation (EC)},
  year = {2018},
  pages = {557--574}
}

@inproceedings{barman2019prop1po,
  author = {Barman, Siddharth and Krishnamurthy, Sanath},
  year = {2019},
  pages = {1748-1755},
  title = {On the Proximity of Markets with Integral Equilibria},
  booktitle = {Proceedings of the 33rd AAAI Conference on Artificial Intelligence (AAAI)}
}

@article{GM23TCS,
  title = {Computing fair and efficient allocations with few utility values},
  journal = {Theoretical Computer Science},
  volume = {962},
  pages = {113932},
  year = {2023},
  author = {Jugal Garg and Aniket Murhekar}
}

@inproceedings{GMQ23chores,
  title = {New Algorithms for the Fair and Efficient Allocation of Indivisible Chores},
  author = {Garg, Jugal and Murhekar, Aniket and Qin, John},
  booktitle = {Proceedings of the Thirty-Second International Joint Conference on Artificial Intelligence (IJCAI)},
  pages = {2710--2718},
  year = {2023}
}

@article{GMQ24wef1po,
  title = {Weighted {EF1} and {PO} Allocations with Few Types of Agents or Chores},
  author = {Garg, Jugal and Murhekar, Aniket and Qin, John},
  journal = {Proceedings of the 33rd International Joint Conference on Artificial Intelligence (IJCAI)},
  year = {2024}
}

@misc{mahara2024polynomialtimealgorithmfairefficient,
  title = {A Polynomial-Time Algorithm for Fair and Efficient Allocation with a Fixed Number of Agents},
  author = {Ryoga Mahara},
  year = {2024},
  eprint = {2411.01810},
  archiveprefix = {arXiv},
  primaryclass = {cs.GT},
  url = {https://arxiv.org/abs/2411.01810}
}
\clearpage
\appendix
\section{Polynomial-Time Implementation of the ER FPTAS}\label{sec:er-bits}

We show how to compute exact optimal flows with logarithmic
costs and supporting prices whose encoding lengths are independent
of the payouts.
Let $\mathcal L$ be the binary encoding length of the rational
market input, and take $\delta$ to be rational. Set $v=n+m$ and
choose $B=O(\mathcal L)$ so that the numerator and denominator
of each $d_{ij}$ are at most $2^B$.

\begin{lemma}[Exact minimum-cost flows]\label{lem:rational-flow}
For rational payouts $q\in Q$, an exact optimal basic flow for
\eqref{eq:flow-value} can be computed in time polynomial in
$\mathcal L$ and the encoding length of $q$.
\end{lemma}
\begin{proof}
We approximate the costs closely enough to preserve every
negative-cost simple residual cycle. A residual arc has cost
$\log d_{ij}$ or $-\log d_{ij}$, so a simple cycle has cost
$\log(A/D)$ for positive integers $A,D\le2^{vB}$. If $A\ne D$, then
\[
 |\log(A/D)|\ge\frac{|A-D|}{\max\{A,D\}}\ge2^{-vB}.
\]
This bound determines how accurately we must approximate the costs.

Compute rational approximations to the costs with error at most
$2^{-vB}/(4v)$. For this, write $d_{ij}=2^k u$ with
$1\le u<2$ and use $\log d_{ij}=k\log2+\log u$, evaluating the
logarithms by a rapidly convergent series. This takes polynomial time. Each simple cycle has at
most $v$ arcs, so its cost changes by at most $2^{-vB}/4$ and its
sign is preserved whenever the original cost is nonzero.
We now solve the rational minimum-cost flow problem, choosing an
optimal basic flow. If this flow were not optimal for the original
costs, it would have a negative simple residual cycle, which would remain
negative under the approximated costs, a contradiction.
Finally, the flow constraint matrix is totally unimodular. Clearing
the denominators of the prescribed payouts therefore gives an
integral basic flow, so the returned flow has polynomial encoding
length.
\end{proof}

\begin{lemma}[Exact supporting prices]\label{lem:rational-prices}
Given an optimal flow with positive chore payouts, we can
compute in polynomial time positive rational numbers $w_j,\alpha_i$,
each with $O(vB)$ bits, such that $\alpha_iw_j\le d_{ij}$ for all
$i,j$, with equality whenever $b_{ij}>0$.
\end{lemma}
\begin{proof}
Write $P_j=\log w_j$ and $A_i=-\log\alpha_i$. The required
conditions become
\[
 P_j-A_i\le\log d_{ij},\qquad
 A_i-P_j\le-\log d_{ij}\quad\text{if }b_{ij}>0.
\]
By \cref{lem:flow-duality}, these inequalities are feasible.
Represent each inequality as a directed arc, and add a source
with zero-length arcs to all vertices. Shortest-path distances
then give feasible values for $P_j$ and $A_i$.

Although the distances may be irrational, we only need their
exponentials. Exponentiating replaces path-length sums by products
of the rational numbers $d_{ij}$, $1/d_{ij}$, and one. Thus we
can run Bellman--Ford using multiplication in place of addition
and exact rational comparisons. Feasibility excludes negative
cycles, so each shortest path can be chosen simple and contains
at most $v-1$ arcs other than its source arc. Consequently, each
exponentiated distance is a product of at most $v-1$ such rationals
and has $O(vB)$ bits. The values $w_j=e^{P_j}$ and
$\alpha_i=e^{-A_i}$ therefore satisfy the required conditions
exactly, with an encoding bound independent of $q$.
\end{proof}

By \cref{lem:entropy} and \eqref{eq:scale-rational}, $\tau,p,r$
also have encoding lengths polynomial in $\mathcal L$, independently
of the iteration number. The initial
payouts have polynomial encoding length, so all subsequent payouts
do as well. By \cref{lem:rational-flow}, each flow computation,
rescaling, and agent earnings test therefore takes polynomial time.
Since $\log R$ is polynomially bounded in $\mathcal L$, the
iteration bound proves that \cref{alg:er} is an FPTAS.

\clearpage
\section{The Chore-Swap Framework}\label{app:framework}

We present the algorithm of \citet{GM26} underlying
\cref{thm:friendly-conversion} and sketch why it works.
Let $Y$ be a $\lambda$-EFX-friendly allocation, with partition
$N=N_0\sqcup N_H$, swap chores $j_i$, and residual bundles $R_i$
as in \cref{def:efx-friendly}. Write $J=\{j_i:i\in N_H\}$.
Fix an arbitrary order of the active agents and relabel them so that
$N_H=\{1,\ldots,k\}$. \Cref{alg:chore-swaps} first reallocates $J$
by letting these agents pick their least-disutility remaining chore
in this order. It then performs chore swaps in the same order.
Writing $g_i$ for the chore picked by agent $i$, the crucial
property of this ordering is $d_i(g_i)\le d_i(g_h)$ whenever $i<h$.
This ordering preserves the $\lambda$-EFX guarantees of earlier agents.

\begin{algorithm}[H]
\caption{Chore swaps from an EFX-friendly allocation \citep{GM26}}
\label{alg:chore-swaps}
\begin{algorithmic}[1]
\Require A $\lambda$-EFX-friendly allocation $Y$, with $N_0,N_H,R_i,j_i$
as in \cref{def:efx-friendly}; $N_H=\{1,\ldots,k\}$.
\Ensure A $\lambda$-EFX allocation $X$.
\State $X_i\gets Y_i$ for $i\in N_0$; $X_i\gets R_i$ for $i\in N_H$;
$J'\gets\{j_i:i\in N_H\}$.
\Statex \textit{Phase 1: Reallocate the swap chores.}
\For{$i=1,\ldots,k$}
  \State Choose $g_i\in\argmin_{j\in J'}d_{ij}$.
  \State $X_i\gets R_i\cup\{g_i\}$; $J'\gets J'\setminus\{g_i\}$.
\EndFor
\Statex \textit{Phase 2: Perform chore swaps in the picking order.}
\For{$i=1,\ldots,k$}
  \If{agent $i$ is not $\lambda$-EFX in $X$}
    \State Choose $\ell\in\argmin_{h\in N}d_i(X_h)$.
    \State $(X_i,X_\ell)\gets(R_i\cup X_\ell,\{g_i\})$.
  \EndIf
\EndFor
\State \Return $X$.
\end{algorithmic}
\end{algorithm}

\paragraph{Proof sketch of \cref{thm:friendly-conversion}.}
We explain why each iteration establishes $\lambda$-EFX for the
agent being processed and preserves it for earlier agents.
Every current bundle contains either a chore from $J$ or an entire
original bundle $Y_h$ for some $h\in N_0$: a swap gives one agent
a chore from $J$ and adds a residual bundle to the other agent's
old bundle. Thus \cref{def:efx-friendly} implies that, throughout
Phase 2,
\begin{equation}\label{eq:swap-residual-bound}
 d_i(R_i)\le(\lambda-1)d_i(X_h)
 \qquad(i\in N_H,\ h\in N).
\end{equation}

\emph{Unprocessed agents retain their bundles.}
Inductively, before agent $i$'s turn, every active agent $h\ge i$
still holds $R_h\cup\{g_h\}$. The picking order and friendliness give
\[
 d_i(R_i)+d_i(g_i)\le(\lambda-1)d_i(g_h)+d_i(g_h)
 \le\lambda d_i(X_h).
\]
Hence $i$ is $\lambda$-envy-free towards all unprocessed agents.
If she violates $\lambda$-EFX, a least-disutility bundle must
therefore belong to an agent in $N_0\cup\{1,\ldots,i-1\}$.
Thus no later agent participates in this swap.

\emph{A swap resolves the current agent's envy.}
Let $B=X_\ell$ be the least-disutility bundle before an $(i,\ell)$
swap. Agent $i$ receives $R_i\cup B$, while $\ell$ is EFX with
the singleton $\{g_i\}$.
By \eqref{eq:swap-residual-bound} and the choice of $B$,
$d_i(R_i\cup B)\le\lambda d_i(X_h)$ for every unchanged bundle
$X_h$. To compare with the new singleton, use the violation of
$\lambda$-EFX before the swap and $d_i(R_i)\le(\lambda-1)d_i(g_i)$:
\[
\begin{aligned}
 d_i(R_i\cup B)
 &<d_i(R_i)+\frac{d_i(R_i)+d_i(g_i)}{\lambda}\\
 &\le\left(1+\frac1\lambda\right)(\lambda-1)d_i(g_i)
       +\frac1\lambda d_i(g_i)
 =\lambda d_i(g_i).
\end{aligned}
\]
\emph{Later swaps preserve these guarantees.}
The only potentially problematic bundle created by a later swap
is its singleton $\{g_h\}$: the other new bundle contains a whole
old bundle. An earlier agent $i$ who now holds a singleton is EFX.
Otherwise, if she initiated a swap, the bound just proved gives
$d_i(X_i)\le\lambda d_i(g_i)\le\lambda d_i(g_h)$.
If she skipped her turn, the same comparison follows from
$X_i=R_i\cup\{g_i\}$, friendliness, and the picking order.
An inactive agent either retains her original bundle, which is
bounded by $\lambda d_i(g_h)$ by friendliness, or holds a singleton.
These observations show inductively that all inactive and processed
agents remain $\lambda$-EFX.

After the last iteration, all agents are $\lambda$-EFX. Each active
agent initiates at most one swap, and each iteration takes polynomial
time. Thus the algorithm uses at most $|N_H|\le n$ swaps and runs
in polynomial time. The full inductive proof appears in \citet{GM26}.

\end{document}